\documentclass[11pt]{article}
\pdfoutput=1
\usepackage{jcapmod}
\usepackage{booktabs}
\usepackage[english]{babel}
\usepackage{amsmath, amssymb, amsfonts, amsbsy, amstext, amsthm}
\usepackage{graphicx}
\usepackage{exscale}
\usepackage[makeroom]{cancel}
\usepackage{soul}
\usepackage{mathtools}
\usepackage{multirow}
\usepackage{caption}
\usepackage{subcaption}
\usepackage{hyperref}

\usepackage{aas_macros}

\allowdisplaybreaks[1]

\numberwithin{equation}{section}

\def\beq{\begin{equation}}
\def\eeq{\end{equation}}
\def\bea{\begin{eqnarray}}
\def\eea{\end{eqnarray}}

\def\q{{\vec{q}\mkern2mu\vphantom{q}}}
\def\x{\vec{x}\mkern2mu\vphantom{x}}
\def\y{\vec{y}\mkern2mu\vphantom{y}}
\def\r{\vec{r}\mkern2mu\vphantom{r}}

\def\xp{\vec{x}\mkern2mu\vphantom{x}'}

\def\planck{{\it Planck} }

\def\Nf{N_{\rm eff}}
\def\Neff{N_{\rm eff}}

\def\To{T^{\rm obs}}
\def\Qo{Q^{\rm obs}}
\def\Uo{U^{\rm obs}}
\def\Eo{E^{\rm obs}}
\def\Bo{B^{\rm obs}}
\def\ao{\alpha^{\rm obs}}
\def\Co{C^{\rm obs}}
\def\Ccr{C^{\rm cross}}
\def\a{\alpha}

\def\Tn{T^{\rm N}}

\def\En{E^{\rm N}}
\def\Bn{B^{\rm N}}
\def\Td{T^{\rm d}}
\def\Qd{Q^{\rm d}}
\def\Ud{U^{\rm d}}
\def\Ed{E^{\rm d}}
\def\Bd{B^{\rm d}}

\def\hP{h^{(P)}}
\def\bhP{\bar{h}^{(P)}}

\def\va{\vec{ \alpha}}
\def\vao{\vec{ \alpha}\,{}^{\rm obs}}
\def\vl{{\vec{\ell}\mkern2mu\vphantom{\ell}}}
\def\vlp{\vec{\ell}\mkern2mu\vphantom{\ell}'}
\def\ml{{L}}
\def\vL{{\vec{L}\mkern2mu\vphantom{L}}}

\DeclareRobustCommand{\SkipTocEntry}[4]{}

\newcommand{\AVE}[1]{\textcolor{red}{#1}}

\begin{document}

\pagenumbering{roman}
\begin{titlepage}
\baselineskip=15.5pt \thispagestyle{empty}

\bigskip\

\vspace{1cm}
\begin{center}

{\fontsize{20.74}{24}\selectfont \sffamily \bfseries  CMB Delensing Beyond the B Modes}

\end{center}
\vspace{0.2cm}
\begin{center}
{\fontsize{12}{30}\selectfont  Daniel Green,$^{\bigstar,\clubsuit}$ Joel Meyers$^{\clubsuit}$, and Alexander van Engelen $^{\clubsuit}$}
\end{center}

\begin{center}

\textsl{$^\bigstar$ University of California, Berkeley, California 94720, USA}
\vskip 7pt

\textsl{$^\clubsuit$ Canadian Institute for Theoretical Astrophysics, Toronto, ON M5S 3H8, Canada}
\vskip 7pt

\end{center}

\vspace{1.2cm}
\hrule \vspace{0.3cm}
\noindent {\sffamily \bfseries Abstract} \\[0.1cm]

Gravitational lensing by large-scale structure significantly impacts observations of the cosmic microwave background (CMB): it smooths the acoustic peaks in temperature and $E$-mode polarization power spectra, correlating previously uncorrelated modes; and it converts $E$-mode polarization into $B$-mode polarization. The act of measuring and removing the effect of lensing from CMB maps, or delensing, has been well studied in the context of $B$ modes, but little attention has been given to the delensing of the temperature and $E$ modes.  In this paper, we model the expected delensed $T$ and $E$ power spectra to all  orders in the lensing potential, demonstrating the sharpening of the acoustic peaks and a significant reduction in lens-induced power spectrum covariances.  We then perform cosmological forecasts, demonstrating that delensing will yield improved sensitivity to parameters with upcoming surveys.  We highlight the breaking of the degeneracy between the effective number of neutrino species and primordial helium fraction as a concrete application.  We also show that delensing increases cosmological information as long as the measured lensing reconstruction is included in the analysis. We conclude that with future data, delensing will be crucial not only for primordial $B$-mode science but for a range of other observables as well.

\vskip 10pt
\hrule

\vspace{0.6cm}
\end{titlepage}

\thispagestyle{empty}
\setcounter{page}{2}
\tableofcontents

\clearpage
\pagenumbering{arabic}
\setcounter{page}{1}

\section{Introduction}
\label{sec:intro}

Weak gravitational lensing of the cosmic microwave background (CMB)~\cite{blanchard87, bernardeau97, zaldarriaga98, lewis06} is now a highly significant feature, seen in  both the power spectra~\cite{calabrese08, reichardt09, keisler11, sievers13, story13, planck13parameters,planck15parameters} and the higher-order statistics~\cite{smith07, hirata08, Das:2011ak, vanEngelen:2012va, polarbear14, story15, planck15lensing}.  Depending on the question of interest, CMB lensing can be either a nuisance or a tool.  For instance, the sum of the neutrino masses can be measured from the reconstruction of the lensing potential~\cite{Kaplinghat:2003bh}.  For some other parameters, error bars improve when the effect of lensing is removed from the power spectra (delensing).  Delensing the $B$-mode polarization to search for primordial gravitational waves is one example that has been studied in great detail~\cite{Knox:2002pe, Kesden:2002ku, Seljak:2003pn, Smith2010, Sherwin:2015baa}, but delensing is a more broadly useful tool that has been explored to a much smaller extent in temperature ($T$) and $E$-mode polarization ($E$).  Recently the delensing of the small-scale temperature field was demonstrated on \planck CMB data using  \planck maps of the cosmic infrared background~\cite{Larsen:2016wpa}.  In this paper we present a computation of the delensed small-scale CMB power spectra to all orders in the lensing potential, calculate the associated covariance matrices of delensed power spectra, and forecast  parameter constraints from upcoming CMB surveys when analyzing the delensed spectra.

\begin{figure}[t!]
\begin{center}
\includegraphics[width=0.75\textwidth]{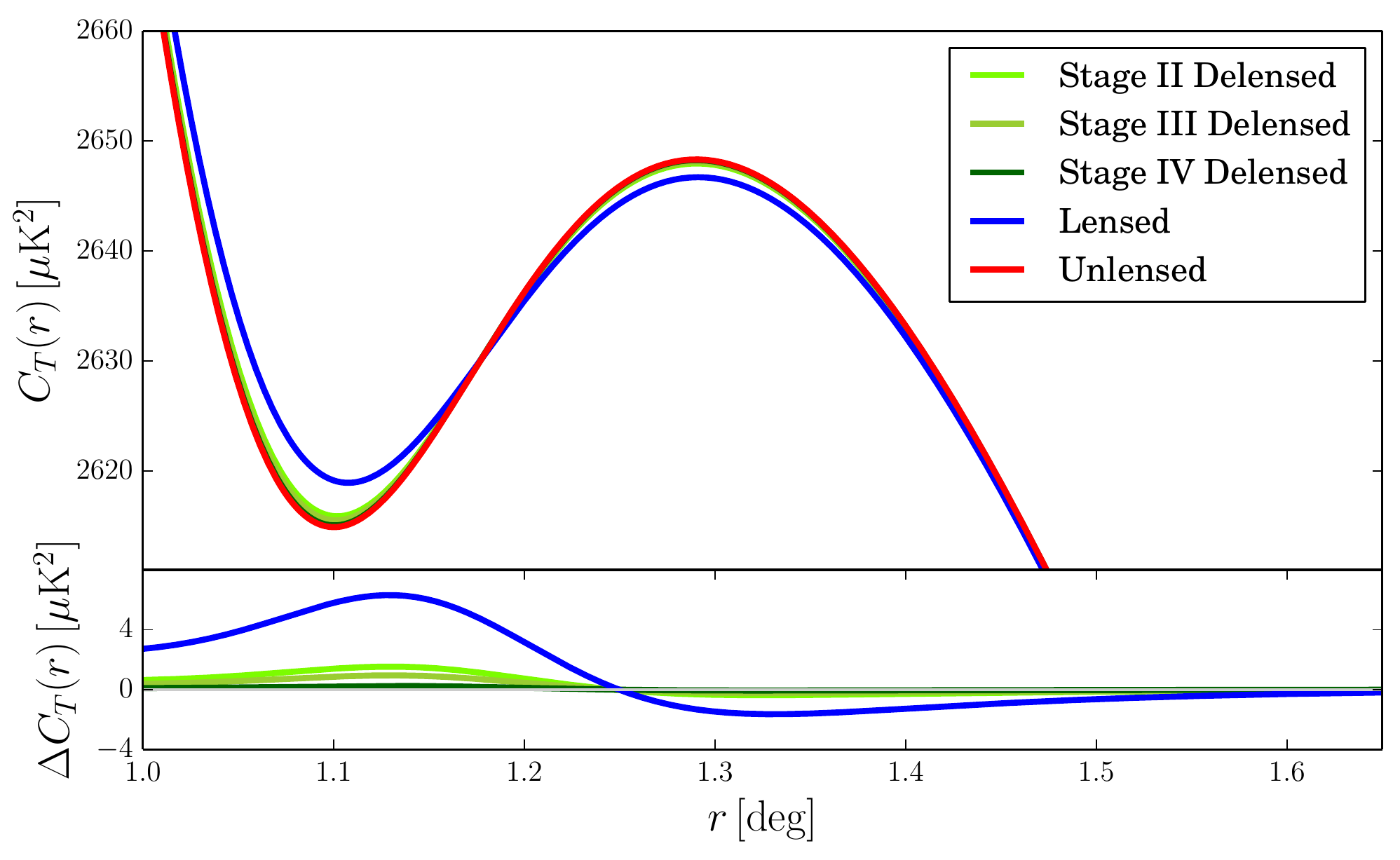}
\caption{The effect of lensing and delensing on the temperature two-point correlation function, $C_T(r)$.  The top panel shows the lensed and  unlensed curves, as well as the delensed curves for various experimental noise configurations using the tools developed in this work.  Specifically, the Stage II, III, and IV experiments contain noise levels of 10, 5, and 1 $\mu$K-arcmin respectively.
 The bottom panel shows the change relative to the unlensed correlation function.  We see that lensing smoothes the BAO feature in the CMB and is restored by delensing, much like what is done with BAO-reconstruction at lower redshifts~\cite{Eisenstein:2006nk}.}
\label{fig:BAO_analogy}
\end{center}
\end{figure}

One of the motivations for studying delensing in this regime is for future measurements of the effective number of neutrino species, $\Neff$.  Free streaming radiation, such as neutrinos, is known to induce a phase shift in the acoustic peaks of the primary CMB~\cite{Bashinsky:2003tk,Follin:2015hya,Baumann:2015rya}.  Lensing is known to  smooth the acoustic peaks~\cite{Seljak:1995ve, Zaldarriaga:1998ar} which, in turn, reduces the accuracy of the measurements of the peak locations.  In fact, the benefit of delensing is quite analogous to BAO reconstruction~\cite{Eisenstein:2006nk}, as illustrated in Figure~\ref{fig:BAO_analogy}.  For this reason, forecasts for future CMB experiments show that unlensed spectra lead to better measurements of $\Neff$~\cite{Baumann:2015rya}.   In reality, delensing is an imperfect procedure and therefore any proper treatment of forecasting or analysis should predict the delensed, rather than unlensed, spectra.  

One of the great technical simplifications of CMB lensing is that the process is local in the observed direction.  In the flat sky limit,
\beq
\tilde T(\x) = T(\x+\va(\x))  \simeq T(\x) + \va(\x)\cdot \vec \nabla T(\x) + \ldots \ .
\eeq
where $\tilde T$ ($T$) is the lensed (unlensed) temperature map, $\vec \alpha = \vec \nabla \phi$ is the deflection angle, and $\phi$ is the lensing potential.  Given an observed temperature map ($\To$) and lensing map ($\vao$), we can certainly imagine a perturbative approach to delensing where
\beq
\Td(\x) \approx \To(\x)  - \vao \cdot \vec \nabla \To(\x) \ ,
\eeq
where $\Td$ is the delensed temperature map.

In practice, modeling lensing of the CMB power spectra requires more accuracy than the simple perturbative description.  Fortunately, $\phi$ is Gaussian to good approximation, which makes an all-orders description of the lensed spectra calculable~\cite{Seljak:1995ve, Zaldarriaga:1998ar}.  One would therefore expect that delensing could be treated by a similar all-orders procedure to predict the delensed power spectra (see also~\cite{Larsen:2016wpa} for a related discussion), given a non-perturbative description of the method for delensing.

In this paper, we will provide an all-orders description of delensing for both temperature and polarization.   We will first describe a non-perturbative approach to delensing that reproduces the unlensed CMB in the limit of no noise.  This procedure is naturally generalized to account for the noise in the temperature, polarization, and lensing maps.  We are careful to use filtered maps as part of the delensing procedure, which we show is necessary for improving parameters constraints.  In principle, one can then produce the all-orders delensed spectra.  In practice, the exact expressions are difficult to calculate due to the non-local relationship between the observed data and the true location of the underlying lenses.  Fortunately, on the scales of interest, the lensing potential varies slowly compared to the CMB maps and these non-local effects can be neglected or included in a perturbative expansion.  This will allow us to provide simple expressions for the delensed power spectra that we also implement numerically.  

The most immediate application of these all-orders results is for forecasting future CMB experiments.  We include forecasts covering a range of possible experimental configurations to illustrate the impact of delensing on $\Neff$ and other cosmological parameters.  Our goal is to understand to what degree forecasts using unlensed spectra are achievable given realistic noise levels in the lensing map.  This is especially important for forecasts of $\Neff$ for CMB Stage IV, which are tantalizingly close to the theoretical threshold of $\Delta \Neff = 0.027$ (see e.g.~\cite{Brust:2013xpv,Salvio:2013iaa,Kawasaki:2015ofa,Chacko:2015noa,Adshead:2016xxj,Baumann:2016wac} for discussion).  We will also show that delensing reduces the covariance between the lensing power spectrum and the observed temperature and polarization spectra. Proper forecasting must thus account for both the delensed spectra and covariance matrix~\cite{BenoitLevy:2012va,Schmittfull:2013uea}.  

This paper is organized as follows.  In Section~\ref{sec:theory}, we present the theoretical framework for computing the delensed CMB spectra.  We apply these results in Section~\ref{sec:sims} to show the numerically computed spectra and covariance matrices.  In Section~\ref{sec:params}, we use these results in forecasts for future CMB experiments.  We highlight the impact of delensing by comparing forecasts with lensed, unlensed, and delensed spectra.  We conclude in Section~\ref{sec:concl}.

The main text is supplemented by six appendices.  Appendix~\ref{app:grad} explores the validity of our expansion in gradients of $\phi$.  We compute the optimal filters in Appendix~\ref{app:filters} for both the temperature and polarization spectra in various limits and explain the choice of filters used in the main text.  Appendix~\ref{app:numpol} gives the expressions for efficient numerical computation of the delensed polarization spectra. In Appendix~\ref{app:covmat}, we show how to calculate the delensed covariance matrix.  In Appendix~\ref{app:info}, we explore the effect of delensing on the Fisher information.  We argue that, as long as the lensing potential is included in the likelihood, one should gain information by delensing.  Appendix~\ref{app:exact} explores an alternate all-orders approach to delensing that is exact in the limit of no noise.  

We will use the following conventions throughout: we define Stage II, III, and IV to be 1 arcmin resolution experiments with 10, 5, and 1 $\mu$K-arcmin temperature noise respectively.  The lensing noise for these experiments is determined assuming the minimum variance quadratic estimator~\cite{Hu:2001kj}, which combines information in the lensed temperature and polarization fields, including the improvement from iterative delensing with the $EB$ reconstruction~\cite{Smith2010}.  We typically show power spectra in terms of ${\mathcal D}_\ell \equiv \ell (\ell+1) C_\ell / (2\pi)$.  We will use $\vl$ to label harmonics of the CMB temperature and polarization but we use $\vL$ for the harmonics of the lensing potential.  
\section{All-Orders Delensing}
\label{sec:theory}

The goal of this section is to present an all-orders theoretical framework for delensing, both in principle and with real data.  Our discussion will focus on temperature at first, but we will also include the generalization to polarization.  We will assume the flat sky limit for simplicity.  Nevertheless, our primary interest is in delensing temperature and $E$-mode polarization where the modes of most interest are well approximated by the flat sky limit.  Previous work aimed at reconstructing the primary (unlensed) CMB  simultaneously with the lensing potential include global maximum-likelihood approaches \cite{hirata03, hirata03b}, local maximum-likelihood approaches \cite{anderes11}, and Bayesian techniques \cite{anderes14}.

Our approach to delensing removes the effect of the lensing directly from the CMB maps, rather than simply deconvolving the estimated lensing deflection power spectrum from the CMB power spectrum.  The realization of the lensing deflection field (including its scatter about the mean) affects the smoothing of the acoustic peaks and is responsible for moving Fisher information from the power spectrum to higher-order statistics.  It is therefore important that delensing is a map-level procedure rather than a deconvolution of the power spectra.  

\subsection{Non-Perturbative Delensing for Real Data}

Lensing is a local process in real space.  Given an unlensed temperature map $T(\x)$ and a lensing field $\va(\x)$, the lensed temperature map is given by
\beq\label{eqn:lens}
\tilde T(\x) = T(\x + \va(\x)) = \int \frac{d^2 \ell}{(2\pi)^2}  e^{i \vl \cdot \big(\x + \va(\x) \big)} T_\vl  \ .
\eeq
Suppose now that we have a perfect measurement of the lensing map (i.e. $\vao = \va$) and would like to reconstruct $T(\x)$.  We can define an exact delensing procedure as
\bea
\Td(\x) = \int d^2 x' J(\x {}') \delta^2(\x-\va(\x') - \x') \tilde  T(\x') \ ,
\eea
where $J(\x) = \det \partial_i (x_j + \alpha_j(\x) )$ is the Jacobian for the change of variables $\y = \x +\va(\x)$ and $\Td$ is our delensed map.  By a change of variables, it is clear that this procedure perfectly inverts the lensing
\beq\label{eqn:exact}
\Td(\x)=  \int d^2 x' J(\x') \delta(\x-\va(\x') - \x') T(\x' +\va(\x')) = \int d^2 y' \delta(\x-\y') T(\y') = T(\x) \ .
\eeq
We see that, in the absence of noise, lensing can be perfectly inverted.

At a practical level, this exact procedure is challenging to implement.  In order to make use of the Jacobian, we need a map of the lensing deflection field, and we also require a map of its gradients at small scales which is typically more contaminated with noise (see Appendix~\ref{app:grad} for more details).   While a simple solution is to neglect the Jacobian in (\ref{eqn:exact}), this approximation introduces an error of ${\cal O}(\vec\nabla \cdot \va)$.

Perhaps the more obvious and convenient approach is to invert lensing locally in the map as follows,
\beq\label{eqn:reverse}
\Td(\x) = \tilde T\big(\x- \va(\x)\big) = T\Big(\x + \va\big(\x-\va(\x)\big) - \va(\x)\Big) =  T\Big(\x + {\cal O}\big((\va \cdot \vec\nabla) \va\big) \Big)\ .
\eeq
One beneficial feature of this approach is that it is very easy to implement given the maps $\tilde T(\x)$ and $\phi(\x)$.  While this procedure does not delens the temperature exactly, it does so up to an error of order ${\cal O}\left( (\va \cdot \vec\nabla) \va\right)$.  This error is acceptably small for our applications as we show explicitly in Appendix~\ref{app:grad}.  We could even iteratively correct this procedure by hand, $\Td(\x) = \tilde T\left(\x - \va(\x) + (\va \cdot \vec\nabla) \va \right)$, to push the error to progressively higher orders in $\va \cdot \vec \nabla$.

\vskip 8pt
In reality, we observe neither the temperature (polarization) nor the lensing potential perfectly.  Given our limited knowledge, we must provide a procedure for delensing an observed map that is well suited for the strengths and limitations of our observations.  We would like any delensing procedure to have the following properties:
\begin{itemize}
\item In the limit where the noise vanishes, delensing should be accurate: $\Td(\x) \approx T(\x)$.
\item We do not add or remove power from the map.  Lensing itself conserves total power and we wish maintain this property of the delensing as well.
\item We should be allowed to filter our maps to minimize the impact\footnote{One often filters maps simply to project out the noisy modes altogether.  Such a procedure does not conserve total power, and we show explicitly in Section~\ref{sec:params} that such a procedure weakens parameter constraints.
  The role of filtering here is to avoid introducing additional noise in the maps from delensing itself, due to noise in $\vao$ and/or $\To$.  See Appendix~\ref{app:filters} for more discussion.} of noisy modes.
\end{itemize}
With real data, we typically want to filter the maps {\it before} delensing to avoid using noisy modes, both in $T$ and $\phi$.  Therefore, 
given an observed temperature map $\To(\x)$, and an observed lensing map $\vao$, our delensed map will be given by
\beq\label{eqn:Tddefinition}
\Td(\x) = \bar h \star \To(\x) + h\star \To(\x - g\star \vao(\x)) 
\eeq
where
\beq
a \star b(\x) = \int d^2 x \, a(\x-{\x}') b(\x') = \int \frac{d^2 \ell}{(2\pi)^2} e^{i \vl \cdot \x} a_\vl  \ b_\vl \ .
\eeq
The functions $\bar h$, $h$, and $g$ are filters that we will discuss shortly.  If we want to satisfy the above requirements, we can see that $\bar h$ must be determined by $g$ and $h$. In particular, there is a constraint imposed on the filters to ensure that we not add or remove power $\langle \Td(0)^2 \rangle = \langle \To(0)^2 \rangle$.  We will determine the explicit expression for of $\bar h$ once we discuss the power spectrum of $\Td$ for general filters\footnote{We discuss optimal and near-optimal filtering schemes in Appendix~\ref{app:filters}.  However, the delensing procedure should make sense independent of the precise choice of filters.}.  In the limit of no noise and $g, h \to 1$, we must impose a constraint $\bar h \to 0$ in order to reproduce the procedure in Equation~(\ref{eqn:reverse}).  An alternative procedure that uses Equation~(\ref{eqn:exact}) as the starting point is discussed in Appendix~\ref{app:exact}.

A similar delensing procedure was recently discussed in~\cite{Larsen:2016wpa}, applied to \planck\ data.  The approach taken there is equivalent to Equation~\ref{eqn:Tddefinition} with $\bar h = 0$ and $h = 1$.  This choice conserves total power and matches our procedure in the limit of no noise.  Including $h\neq 1$ and $\bar h \neq 0$ is important for minimizing the noise induced in the delensed maps, as shown in Appendix~\ref{app:noise}.  Furthermore, we will show in Section~\ref{sec:params} that filtering with $\bar h = 0$ produces significantly weaker constraints on cosmological parameters than when we allow $\bar h \neq 0$ for noisy modes.  

The expressions for the delensed maps are easier to work with in terms of harmonics,
\beq
\Td_\vl = \bar h_\vl \To_\vl + \int d^2 x  \frac{d^2 \ell_1 }{(2 \pi)^2} h_{\vl_1} \To_{\vl_1} e^{-i \vl\cdot \x} e^{i \vl_1 \cdot (\x-g \star \va(\x) )} \ .
\eeq
From here on, we will assume isotropic noise, and therefore isotropic filters, so we will take $h_\vl = h_\ell$ and $\bar h_\vl = \bar h_\ell$.
We can formally write the delensed $C_\ell$ as
\bea
\langle \Td_\vl \Td_{\vlp} \rangle &=& (2\pi)^2 \delta(\vl+\vlp) |\bar h_\ell|^2 C_\ell^{\rm obs}+\Bigg[ \int d^2 x \frac{ d^2 \ell_1}{(2\pi)^2} e^{-i (\vl-\vl_1) \cdot \x} \times \nonumber \\
&& \qquad \qquad  \bigg(  \bar h_{\ell'} h_{\ell_1} \langle  e^{-i \vl_1 \cdot g\star \vao(\x)} \To_{\vl_1} \To_{\vlp} \rangle \bigg) + \{ \vl \leftrightarrow \vlp \} \Bigg] \nonumber \\
&& + \int d^2 x'  d^2 x'' \frac{ d^2 \ell_1d^2 \ell_2}{(2\pi)^4}  e^{-i (\vl-\vl_1) \cdot \x}  e^{-i (\vlp-\vl_2) \cdot \xp} h_{\ell_1} h_{\ell_2 } \times \nonumber  \\
&& \qquad  \qquad  \langle  e^{-i \vl_1 \cdot g\star \vao(\x)} e^{-i \vl_2 \cdot g\star \vao(\x')}  \To_{\vl_1} \To_{\vl_2} \rangle \label{eqn:noapproxTT} \ .
\eea
In principle, this defines the all-orders result, if one can evaluate all of the correlation functions exactly and perform the integrals.  In the next section, we will discuss several approximations that will simplify the calculations and allow for simpler analytic results that can be efficiently computed numerically.  We can also systematically improve these (or any other) approximations by treating them as a perturbative expansion of this equation.

While these results hold for any choice of the $g_L$, $h_\ell$, and $\bar h_\ell$ filters, one should optimize these choices to minimize the impact of noise.  We will discuss this optimization in Appendix~\ref{app:filters} but we will ultimately use signal-to-noise filtering of the form
\bea
g_L = \frac{C_L^{\phi \phi}}{C_L^{\phi \phi, {\rm obs}}} \qquad \qquad h_\ell = \frac{\tilde C_\ell^{TT}}{C_\ell^{TT, {\rm obs}}} \ ,
\eea
where $\tilde C^{TT}_\ell$ is the lensed spectrum.  We determine $\bar h_\ell$ in terms of $h_\ell$ and $g_L$ from the conservation of total power in Appendix~\ref{app:totalpowerT}.  We will also need to define a filter for polarization, $h^P_\ell$, which is similarly filtered in terms of $C_\ell^{EE}$. These choices are optimal in certain limits, but are also easy to implement and have simple interpretations.

\subsection{Approximate Delensed Power Spectra}

Two well-motivated approximations that will greatly simplify calculations are (1) small lensing gradients, $(\va \cdot  \vec \nabla) \, \va\ll \va$ and (2) Gaussianity of the lensing potential.  These two approximations alone allow for simple all-orders expressions for the delensed power spectra.  

Let us start by dropping gradients of $\va$.  As described above, in the limit of no noise we can make the approximation $\tilde T(\x - g\star \vao(\x)) \approx T(\x+\va-g\star \vao(\x))$ up to errors that we can safely ignore (see Appendix~\ref{app:grad} for details).  However, we are still filtering the maps before delensing, which makes it more challenging to implement this approximation.  Specifically, we have
\bea
 h \star \To(\x - g \star \vao(\x)) &=&\int d^2 x'  \frac{d^2 \ell_1 }{(2 \pi)^2} h_{\ell_1}e^{i \vl_1 \cdot (\x-\xp-g \star \vao(\x) )}  \To(\x')  \\
&=& \int  \frac{d^2 \ell_1 }{(2 \pi)^2}  h_{\ell_1} e^{i \vl_1 \cdot (\x-g \star \vao(\x) )} \Big[ \Tn_{\vl_1} +e^{-i\vl_1 \cdot \x } \int d^2 x' \frac{d^2 \ell_2 }{(2 \pi)^2} e^{ i \vl_2 \cdot (\xp+\va(\x') )} T_{\vl_2}\Big] \ ,   \nonumber 
\eea 
where we used $\To = \tilde T+ T^{\rm N}$.  For a completely general filter, $h_\ell$, this expression can be challenging to work with.  In practice, for the range of interest, the filters change very slowly.  For $\ell_1, \ell_2$ that dominate these integrals, we expect that $h_{\ell_1} \simeq h_{\ell_2}$, so we can perform the integral over $\ell_1$ (and relabel $\ell_2 \to \ell_1$) to find   
\bea
 h \star \To(\x - g \star \vao(\x))&\approx& \int  \frac{d^2 \ell_1 }{(2 \pi)^2}  h_{\ell_1}  \Big[ e^{i \vl_1 \cdot (\x-g \star \vao(\x) )} \Tn_{\vl_1} +e^{ i \vl_1 \cdot (\x - g \star \vao(\x) +\va(\x) )} T_{\vl_1}\Big] \ ,   \nonumber \\
  \bar h \star \To(\x)&\approx& \int  \frac{d^2 \ell_1 }{(2 \pi)^2}  \bar h_{\ell_1}  \Big[ e^{i \vl_1 \cdot \x} \Tn_{\vl_1} +e^{ i \vl_1 \cdot (\x +\va(\x) )} T_{\vl_1}\Big] \ .
\eea
In position space, we are assuming that $h(\x)$  is highly localized (i.e. approximately a delta function) compared to the scales over which $\va(\x)$ varies.  In $\ell$-space, this means that $h_\ell \simeq h_{\ell \pm 100}$ since the lensing power spectrum peaks at large angular scales $L\lesssim 100$.

Putting this all together and transforming to harmonics gives
\bea
\Td_\vl &=& \bar h_\ell \Tn_\vl + \int d^2 x  \frac{d^2 \ell_1 }{(2 \pi)^2}  \Bigg[ h_{\ell_1} e^{-i \vl \cdot \x}  \Big( e^{i \vl_1 \cdot (\x-g \star \vao(\x) )} \Tn_{\vl_1} +e^{ i \vl_1 \cdot (\x - g \star \vao(\x) +\va(\x) )} T_{\vl_1}\Big) \nonumber \\
&& \hskip 120pt + \bar h_{\ell_1} e^{-i \vl \cdot \x }e^{i \vl_1 \cdot (\x+ \va(\x))} T_{\vl_1} \Bigg] \ . \label{eqn:ttresult}
\eea
The noise and unlensed temperature are independent of the lensing potential, and therefore the power spectrum becomes
\bea
\left\langle \Td_\vl \Td_{\vlp} \right\rangle &=& (2\pi)^2 \delta(\vl+\vlp) |\bar h_\ell|^2 C_\ell^{\rm obs}+\Bigg[ \int d^2 x \, \bar h_{\ell'} h_{-\ell'} C^{\rm N}_{\ell'} e^{-i (\vl+\vlp) \cdot \x} \left\langle e^{-i \vlp \cdot g\star \vao(\x)}\right\rangle + \{ \vl \leftrightarrow \vlp \} \Bigg] \nonumber \\
&& +  \int d^2 x d^2 x' \frac{d^2 \ell_1}{(2\pi)^2} \, \bar h_{\ell_1} h_{-\ell_1} C_{\ell_1}  \times \nonumber \\
&& \qquad  \qquad \Bigg[ e^{-i \vl \cdot \x -i \vlp \cdot \xp +i\vl_1\cdot (\x-\xp) } \left\langle e^{i \vl_1\cdot (\va(\x)-  g\star \vao(\x))}  e^{-i \vl_1\cdot \va(\x')}\right\rangle  + \{ \vl \leftrightarrow \vlp \} \Bigg] \nonumber \\
&& +  \int d^2 x d^2 x' \frac{d^2 \ell_1}{(2\pi)^2} \, | h_{\ell_1}|^2 C_{\ell_1}  \times \nonumber \\
&& \qquad  \qquad \Bigg[ e^{-i \vl \cdot \x -i \vlp \cdot \xp +i\vl_1\cdot (\x-\xp) } \left\langle e^{i \vl_1\cdot (\va(\x)-  g\star \vao(\x))}  e^{-i \vl_1\cdot (\va(\x')-  g\star \vao(\x'))}\right\rangle \Bigg]  \label{eqn:harmonicsapprox} \ .
\eea
In principle, this reduces the problem to the correlation functions of $\va$ and $\vao$, but because of the exponentials, it will involve an infinite series of terms.

Now we will assume\footnote{The CMB lensing field is expected to be mostly Gaussian because the matter fluctuations that lens the CMB are mainly at high redshift $z\sim 2$ and on large scales $k \sim 10^2$Mpc$^{-1}$, where the perturbations are well-described by linear theory .  This is supported by simulations~\cite{Antolini:2013ika, liu16} and analytical calculations ~\cite{Bohm:2016gzt, namikawa16, lewis16}.  Non-Gaussianity from post-Born corrections should also be small \cite{pratten16}.  Non-Gaussian contributions may also be included as perturbative corrections to our Gaussian approximation.} that $\va$ is Gaussian.  We can then evaluate the correlation functions using
\beq\label{eq:expcorr}
\langle e^{i y} \rangle = \int_{-\infty}^{\infty}  dy e^{iy} \frac{1}{\sqrt{2\pi \sigma^2}} e^{-y^2/2\sigma^2}  = e^{-\langle y^2 \rangle /2} \ ,
\eeq
which holds for any Gaussian random variable $y$.  

We can now reduce the problem to computing the two-point correlation functions of $\va$ and $\vao$ in terms of $C^{\phi \phi}_\ell$ and $C^{\phi \phi, {\rm obs}}_\ell$ where $\va = \vec \nabla \phi$.  This boils down to evaluating
\beq
\left\langle \alpha^X_i(\x) \alpha^Y_j(\x') \right\rangle = \frac{1}{2} \delta_{ij}  C^Z_0(r) - (\hat r^i \hat r^j - \frac{1}{2} \delta^{ij}) C^Z_2(r)
\eeq
where $X,Y = \{ \, , {\rm obs} \}$ and $Z = \{ \,  , {\rm cross} , {\rm obs} \}$ for $\{ X=Y\neq {\rm obs} \, , X\neq Y, X=Y = {\rm obs} \}$ respectively.  With these conventions, we will define the filter to be ${\bf g}_L^Z = \{1, g_L , |g_L|^2 \}$ and lensing power $C_L^{\phi \phi, Z} = \{ C^{\phi \phi}_L , C_L^{\phi \phi},C_L^{\phi \phi, {\rm obs}} \}$.  We have assumed isotropic reconstruction noise so that  $g_{\vL} = g_L$.  In terms of these quantities, our correlation functions are given by
\bea
C^Z_0(r)\equiv  \left\langle \va(\x) \cdot \va(\x') \right\rangle &=& \int \frac{d^2 L }{(2\pi)^2} L^2 {\bf g}^Z_L C^{\phi\phi, Z}_L e^{i \vL \cdot \r} \nonumber \\
&= &\int \frac{ d L }{2\pi} L^3 {\bf g}^Z_L C^{\phi\phi, Z}_L J_0( L r) \ ,
\eea
and
\bea
C^Z_2(r)\equiv  -2 (\hat r^i \hat r^j - \frac{1}{2} \delta^{ij})\left\langle \alpha_i(\x) \alpha_j(\x') \right\rangle &=& - \int \frac{d^2 L }{(2\pi)^2} L^2 \cos 2 \varphi \, {\bf g}^Z_L C^{\phi\phi,Z}_L e^{i \vL \cdot \r}  \nonumber \\
&=&  \int \frac{ d L }{2\pi} L^3 J_2(L r) {\bf g}^Z_L C^{\phi\phi,Z}_L  \ ,
\eea
where $\vL \cdot \hat r = L \cos\varphi$.  In practice, these quantities are all very similar, and they differ only in the sense that we are filtering the observed map (but not the underlying lensing field $\va(\x)$) and that the noise in our measurement only affects the power spectrum of $\vao$.  The notation in terms of $C_{0,2}^Z$ captures that there are two power spectra and one cross-correlation.

Putting this all together, we get
\bea
\left\langle \Td_\vl \Td_{\vlp} \right\rangle &=& (2\pi)^2 \delta(\vl+\vlp)\Bigg[ |\bar h_\ell|^2 C_\ell^{\rm obs}+ ( h_{\ell} \bar h_{-\ell}+h_{-\ell} \bar h_{\ell}) e^{-\tfrac{1}{4} \ell^2 \Co_0(0)} C_{\ell}^{\rm N} \label{eqn:finalTd} \\
&&+  \int d^2 r \frac{d^2 \ell_1 }{(2\pi)^2} |h_{\ell_1}|^2 C_{\ell_1}^{\rm N} e^{-i (\vl - \vl_1)\cdot \r} e^{-\frac{\ell_1^2}{2} (\Co_0(0)-\Co_0(r)+ \cos 2\varphi_1\Co_2(r))} \nonumber \\
&&+ \int d^2 r \frac{d^2 \ell_1 }{(2\pi)^2}   C_{\ell_1} e^{-i (\vl - \vl_1)\cdot \r} e^{-\frac{\ell^2_1}{2} (C_0(0)-C_0(r)+ \cos 2\varphi_1 C_2(r))} \nonumber \\
&&\times \Bigg( (h_{\ell_1} \bar h_{-\ell_1}+h_{-\ell_1} \bar h_{\ell_1}) e^{-\frac{\ell_1^2}{4} \Co_0(0) +\frac{\ell_1^2}{2} (\Ccr_0(0)  - \Ccr_0(r))+\tfrac{\ell_1^2}{2}  \cos2 \varphi_{1} \Ccr_2(r))}  \nonumber \\
&& + |h_{\ell_1}|^2 e^{-\frac{\ell_1^2}{2} (\Co_0(0)-\Co_0(r) + \cos 2\varphi_1 \Co_2(r) )+\ell_1^2(\Ccr_0(0) - \Ccr_0(r))+  \ell_1^2 \cos2 \varphi_{1}  \Ccr_2(r)} \Bigg)\Bigg] \ .\nonumber
\eea
where $\hat \ell_1 \cdot \hat r =  \cos \varphi_1$.  When computing spectra in later sections, we will often consider the case where cross-correlations are used to effectively set $C^N_\ell = 0$ in this expression.  However, when computing the covariance matrices the noise terms cannot be removed. 

From these expressions, we can also determine the $\bar h_\ell$ that conserves total power.  This is derived in Appendix~\ref{app:totalpowerT} and give the result
\beq
\bar h_{\ell} = \sqrt{1- h_\ell^2 \left(1-e^{- \frac{\ell^2}{2} \Co_0(0)}\right)} - h_\ell e^{- \frac{\ell^2}{4} \Co_0(0)} \ .  \label{eq:hbar}
\eeq
To zeroth order in $\Co_0(0)$, this expression is $\bar h_\ell \approx 1 - h_\ell$.

\subsection{Polarization}

Lensing of the polarization field is quite familiar when written in terms of $Q$ and $U$, 
\beq
\big[ Q \pm i U\big] (\x) = - \int \frac{d^2 \ell}{(2\pi)^2} \big[ E \pm i B\big]_\vl \,  e^{\pm 2 i \varphi_\vl} e^{i \vl \cdot \x} \ ,
\eeq
where the factor $e^{\pm 2 i \varphi_\vl}$ converts between the fixed reference basis in real space and the natural basis in harmonic space.  Like temperature, lensing moves the points on this map via
\beq
\big[ \tilde Q \pm i \tilde U\big] (\x) = \big[  Q \pm i U\big] (\x+\va(\x)) = - \int \frac{d^2 \ell}{(2\pi)^2} \big[ E \pm i B\big]_\vl \, e^{\pm 2 i \varphi_\vl} e^{i \vl \cdot (\x+\va(\x))} \ .
\eeq 
The all-orders delensing procedure at this level is therefore to take the $Q$ and $U$ maps and apply the same procedure as we did for temperature:
\bea
\big[ \Qd\pm i  \Ud \big] (\x)  =  \bhP \star \big[\Qo\pm i \Uo \big] (\x)  + \hP \star \big[\Qo\pm i \Uo \big] (\x-g\star \vao(\x))  \ .
\eea
We note that, a priori,  $\bhP$ and $\hP$ are not required to be real functions.  Yet, in practice, delensing is a procedure we apply to the $Q$ and $U$ maps, and if the noise is isotropic then there should be no distinction between the filtering of $Q$ and $U$ nor should they be mixed under filtering.  As result, we will choose our filters to be real (see Appendix~\ref{app:EBfilter} for further discussion).  

Making the same approximations as we did for temperature, the delensed $E$ and $B$ are given by
\bea
(\Ed \pm i \Bd)_\vl &=& \bhP_{\ell} (\Eo \pm i \Bo)_\vl + \int d^2 x \frac{d^2 \ell_1}{(2\pi)^2} \hP_{\ell_1} e^{\pm 2 i (\varphi_{\vl_1} - \varphi_\vl)}e^{-i (\vl-\vl_1) \cdot \x} \nonumber \\
&&\times \Bigg[  (\En\pm i \Bn)_{\vl_1} e^{ -i \vl_1 \cdot g \star \vao(\x)}+ (E \pm i B)_{\vl_1}  e^{i \vl_1 \cdot (\va(\x) - g \star \vao(\x))}\Bigg] \ .
\eea
Now we can isolate $\Ed$ (or equivalently $\Bd$) as 
\bea
\Ed_\vl &=&\bhP_{\ell} \Eo_\vl + \int d^2 x \frac{d^2 \ell_1}{(2\pi)^2} \hP_{\ell_1} e^{i (\vl_1-\vl) \cdot \x} \nonumber \\
&&\times \Bigg[  \cos(2 (\varphi_{\vl_1} - \varphi_\vl)) \bigg( \En_{\vl_1}e^{ -i \vl_1 \cdot g \star \vao(\x)}+ E_{\vl_1}  e^{i \vl_1 \cdot (\va(\x) - g \star \vao(\x))} \bigg) \nonumber \\
&&+ \sin(2 (\varphi_{\vl_1} - \varphi_\vl))\Bn_{\vl_1}e^{ -i \vl_1 \cdot g \star \vao(\x)} \Bigg] \ .
\eea
where we have assumed there are no primordial B-modes.  Repeating the same steps as for temperature, we find 
\bea
\langle \Ed_\vl \Ed_{\vlp} \rangle &=& (2\pi)^2 \delta(\vl+\vlp)\Bigg[ \left|\bhP_\ell\right|^2 C^{EE, {\rm obs}}_\ell+ ( \hP_{\ell} \bhP_{-\ell}+\hP_{-\ell} \bhP_{\ell}) e^{-\tfrac{1}{4} \ell^2 \Co_0(0)} C_{\ell}^{EE,\rm{N}} \label{eqn:finalEd} \\
&&+  \int d^2 r \frac{d^2 \ell_1 }{(2\pi)^2} e^{-i (\vl - \vl_1)\cdot \r} e^{-\frac{\ell_1^2}{2} (\Co_0(0)-\Co_0(r)+ \cos 2\varphi_1 \Co_2(r))} \nonumber \\
&&\times \left| \hP_{\ell_1}\right|^2  \Big[ \cos^2(2 (\varphi_{\vl_1} - \varphi_\vl)) C_{\ell_1}^{EE, \rm{N}} +\sin^2(2 (\varphi_{\vl_1} - \varphi_\vl))  C_{\ell_1}^{BB, \rm{N}} \Big] \nonumber \\
&&+ \int d^2 r \frac{d^2 \ell_1 }{(2\pi)^2}   \cos^2(2 (\varphi_{\vl_1} - \varphi_\vl)) C_{\ell_1}^{EE}e^{-i (\vl - \vl_1)\cdot \r} e^{-\frac{\ell_1^2}{2} (C_0(0)-C_0(r)+ \cos 2\varphi_1 C_2(r))} \nonumber \\
&&\times \Bigg( ( \hP_{\ell_1} \bhP_{-\ell_1}+\hP_{-\ell_1} \bhP_{\ell_1})  e^{-\frac{\ell_1^2}{4} \Co_0(0) +\frac{\ell_1^2}{2} \left(\Ccr_0(0)  - \Ccr_0(r)+ \cos2 \varphi_{1} \Ccr_2(r)\right)}  \nonumber \\
&& + \left|\hP_{\ell_1}\right|^2 e^{-\frac{\ell_1^2}{2} \left(\Co_0(0)-\Co_0(r) + \cos 2\varphi_1 \Co_2(r) \right)+\ell_1^2 \left(\Ccr_0(0) - \Ccr_0(r) + \cos2 \varphi_{1}  \Ccr_2(r)\right) } \Bigg)\Bigg] \ , \nonumber
\eea
where we used $\varphi_{-\vl} = \varphi_\vl + \pi$.  Similarly we can compute the $TE$ correlation, which simplifies significantly
\bea
\left\langle \Td_\vl \Ed_{\vlp} \right\rangle &=& (2\pi)^2 \delta(\vl+\vlp)\Bigg[  \bar h_{\ell} \bhP_{-\ell} \tilde C^{TE}_\ell \label{eqn:finalTEd} \\
&&+ \int d^2 r \frac{d^2 \ell_1 }{(2\pi)^2}   \cos(2 (\varphi_{\vl_1} - \varphi_\vl)) C_{\ell_1}^{TE}e^{-i (\vl - \vl_1)\cdot \r} e^{-\frac{\ell_1^2}{2} (C_0(0)-C_0(r)+ \cos 2\varphi_1 C_2(r))} \nonumber \\
&&\times \Bigg( ( h_{\ell_1} \bhP_{-\ell_1}+\hP_{-\ell_1} \bar h_{\ell_1})  e^{-\frac{\ell_1^2}{4} \Co_0(0) +\frac{\ell_1^2}{2} \left(\Ccr_0(0)  - \Ccr_0(r) + \cos2 \varphi_{1} \Ccr_2(r)\right)}  \nonumber \\
&& + h_{\ell_1} \hP_{-\ell_1} e^{-\frac{\ell_1^2}{2} \left(\Co_0(0)-\Co_0(r) + \cos 2\varphi_1 \Co_2(r) \right) + \ell_1^2 \left(\Ccr_0(0) - \Ccr_0(r) + \cos2 \varphi_{1}  \Ccr_2(r)\right) } \Bigg)\Bigg] \ .\nonumber
\eea
The cross-correlation in this case will have some unusual features because we have separate filters acting on $T$ and $E$.  At the map level, this means there are modes where we delens the temperature but not the polarization ($h_\ell \approx 1$ and $\bhP_\ell \approx 1$).  The temperature and polarization information is carried by the same photons yet this delensing procedure can break this physical relationship between the lensing in the two maps.  Nevertheless, our filtering scheme is chosen to minimize the error-bars in our cosmological parameters and some counter-intuitive features arise as a consequence of minimizing the noise.  

One can also apply the same technique to delensing the $B$ modes.  Delensing of the $B$ modes is substantially simplified compared to $T$ and $E$ because the unlensed signal is often assumed to vanish.  Alternate delensing schemes, such as iterative delensing~\cite{Smith2010}, are likely to be effective in this context and can also be used to generate the $\phi$-map.

\vskip 8pt
\noindent{\it Summary:} \hskip 2pt Together with $C^{\phi \phi}_{L}$, the delensed spectra given by Equations (\ref{eqn:finalTd}), (\ref{eqn:finalEd}) and (\ref{eqn:finalTEd}) will form the basis for forecasting presented in Section~\ref{sec:params}.  Forecasts involving the tensor-to-scalar ratio, $r$, should include $C^{{\rm d}, BB}_\ell$ but is beyond the scope of this work. In principle, these same techniques can also be generalized to any $N$-point function or corrected perturbatively to improve accuracy. 

\section{Numeric Spectra and Covariances}
\label{sec:sims}

In the previous section, we gave analytic formulas for the $TT$, $TE$, and $EE$ spectra given the unlensed spectra and the noise power spectra for the CMB and lensing maps (assuming Gaussian, isotropic noise).  Evaluating these expressions must be done numerically but this is straightforward given the existing techniques used to compute the lensed spectra~\cite{lewis06}.  Furthermore, from these expressions, we can also compute the covariance matrix for the delensed spectra.  In this section we will show the numeric calculation of the spectra and covariance matrix.  
\subsection{Spectra}

\begin{figure}[t!]
\begin{center}
\includegraphics[width=\textwidth]{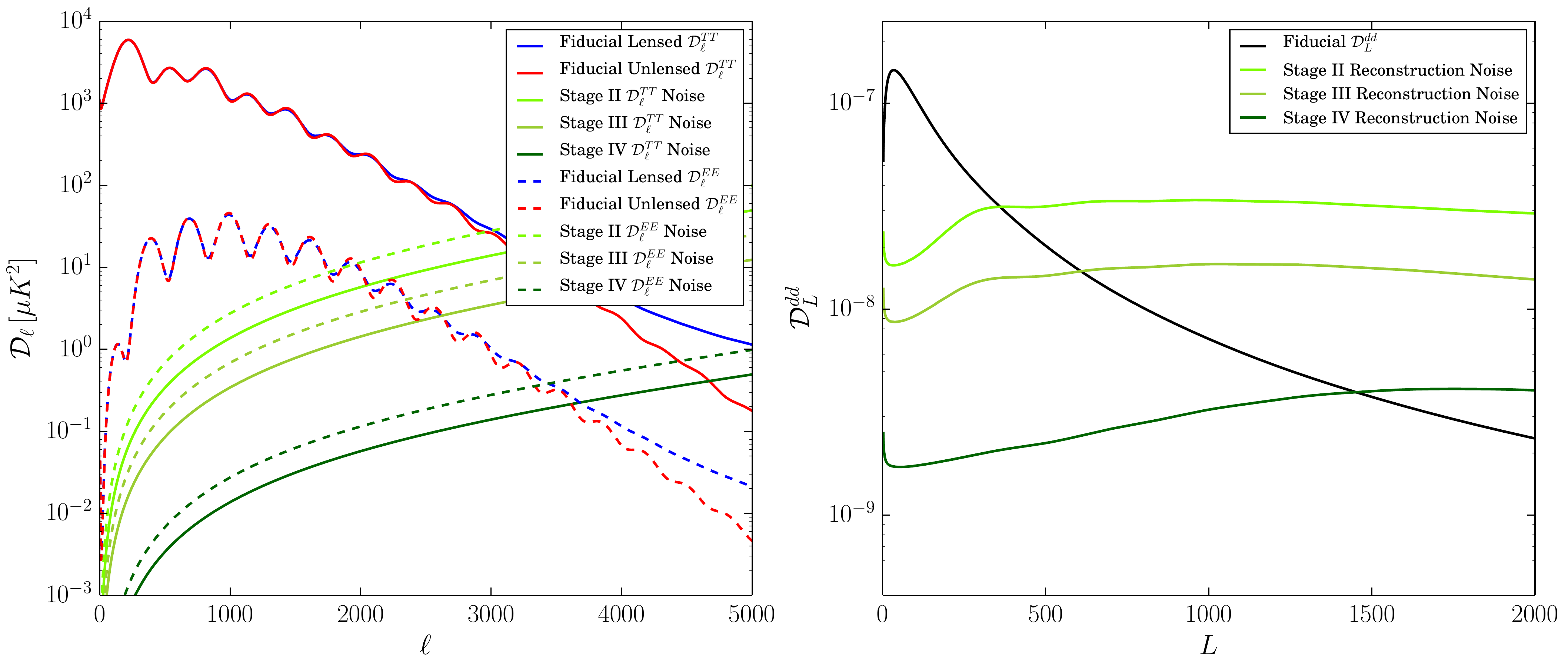}
\caption{{\it Left:} Noise spectra for both $T$ and $E$ for Stage II, III, and IV experiments (green) compared to the lensed (blue) and unlensed (red) $TT$ (solid) and $EE$ (dashed) spectra.  In what follows, the transition region from signal-dominated to noise-dominated is where we expect filtering to play an important role. {\it Right:} Lensing deflection ($d \equiv L \phi$) power spectrum in black, compared to the lensing deflection noise for Stage II, III, and IV experiments in green (details can be found in Section~\ref{sec:forecastmethod}).  We expect that modes that are reconstructed with high signal-to-noise can be removed from CMB maps by delensing.}\label{fig:noise_power_combined}
\end{center}
\end{figure}

The method we use to numerically compute the delensed power spectra is very similar to the procedure for computation of the flat-sky lensed spectra in CAMB \cite{Lewis2000}.  Numerical results are more stable if one computes first the change to the unlensed spectrum due to the combination of lensing and delensing, and then adds the result to the unlensed spectrum.  Starting from Eq.~(\ref{eqn:finalTd}), we expand to first order in $C_2(r)$ and find for the change to the delensed temperature correlation function
\bea
\Delta\xi_T^{\rm d}(r) &=& \int \frac{d\ell}{2\pi} \ell C_\ell \Bigg[ -J_0(\ell r) \label{eqn:delta_xiTd} \\
&& + \left|\bar h_\ell\right|^2 \exp\left[-\frac{\ell^2}{2} (C_0(0)-C_0(r))\right]\left(J_0(\ell r) + \frac{\ell^2}{2} C_2(r) J_2(\ell r)\right)  \nonumber \\
&& + \left(h_{\ell} \bar h_{-\ell}+h_{-\ell} \bar h_{\ell}\right) \exp\left[-\frac{\ell^2}{2} \left((C_0(0)-C_0(r)) - (\Ccr_0(0)-\Ccr_0(r)) + \frac{1}{2}\Co_0(0)\right)\right] \nonumber \\
&& \quad \times \left(J_0(\ell r) + \frac{\ell^2}{2} \left(C_2(r) - \Ccr_2(r)\right) J_2(\ell r)\right) \nonumber \\
&& + \left|h_{\ell}\right|^2 \exp\left[-\frac{\ell^2}{2} \left((C_0(0)-C_0(r)) - 2(\Ccr_0(0)-\Ccr_0(r)) + (\Co_0(0)-\Co_0(r))\right)\right] \nonumber \\
&& \quad \times \left(J_0(\ell r) + \frac{\ell^2}{2} \left(C_2(r) - 2\Ccr_2(r) + \Co_2(r)\right) J_2(\ell r)\right)\Bigg] \ . \nonumber
\eea
One could include higher orders in $C_2(r)$ for improved accuracy, however most scales in the CMB are well approximated with the first-order approach used here~\cite{Challinor:2005jy,lewis06}. These expressions are evaluated using $h_\ell = \tilde C_\ell / (\tilde C_\ell + C^{N}_\ell)$, as described in Appendix~\ref{app:filters}.  These spectra also require the noise for both $\phi$ and $T$ which is shown in Figure~\ref{fig:noise_power_combined}.

The delensed temperature power spectrum is then given by
\beq
C_\ell^{\rm d} = C_\ell + 2\pi \int rdr J_0(\ell r) \Delta\xi_T^{\rm d}(r) \, . \label{eqn:delensed_temp_spec}
\eeq 
Similar remarks apply for the polarization spectra, which are discussed in Appendix~\ref{app:numpol}.

\begin{figure}[t!]
\begin{center}
\includegraphics[width=0.85\textwidth]{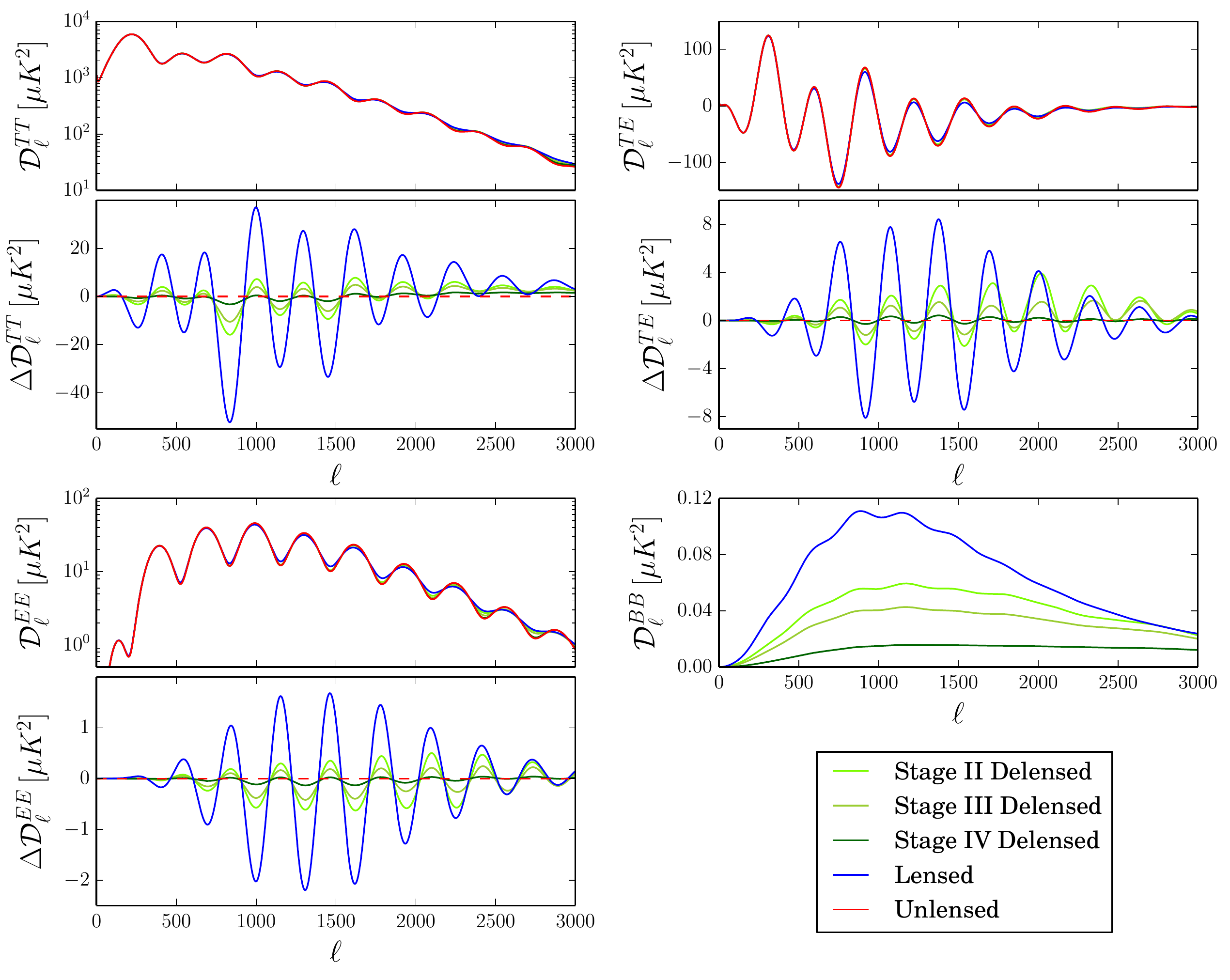}
\caption{Comparison of lensed, unlensed, and delensed $TT$, $TE$, $EE$, and $BB$ power spectra for Stage II, III, and IV experiments using the cosmological parameters listed in Table~\ref{table:cosmo_fiducial}.   For each spectrum, in the lower panel we show $\Delta \mathcal{D}_\ell \equiv \mathcal{D}_\ell - \mathcal{D}_\ell^{\rm unlensed}$ in order to highlight the effect of delensing on the power spectra.  The spectra for an experiment with perfect delensing would lie on the zero line in these lower panels. }
\label{fig:delensed_spectra_square}
\end{center}
\end{figure}

The delensed spectra that result from this calculation are shown in Figure~\ref{fig:delensed_spectra_square}.  Qualitatively, the delensed spectra follow our basic expectations.  At low $\ell$, where the temperature and polarization are measured with high signal-to-noise, our ability to delens is limited only by the lensing reconstruction noise.  At high $\ell$, when the temperature and polarization noise in the left panel of Figure~\ref{fig:noise_power_combined} begins to dominate over the signal, delensing becomes difficult and delensed spectra follow the lensed spectra.  Both statements are consistent with expectations from the form of the filters.  

\begin{figure}[t!]
\begin{center}
\includegraphics[width=0.75\textwidth]{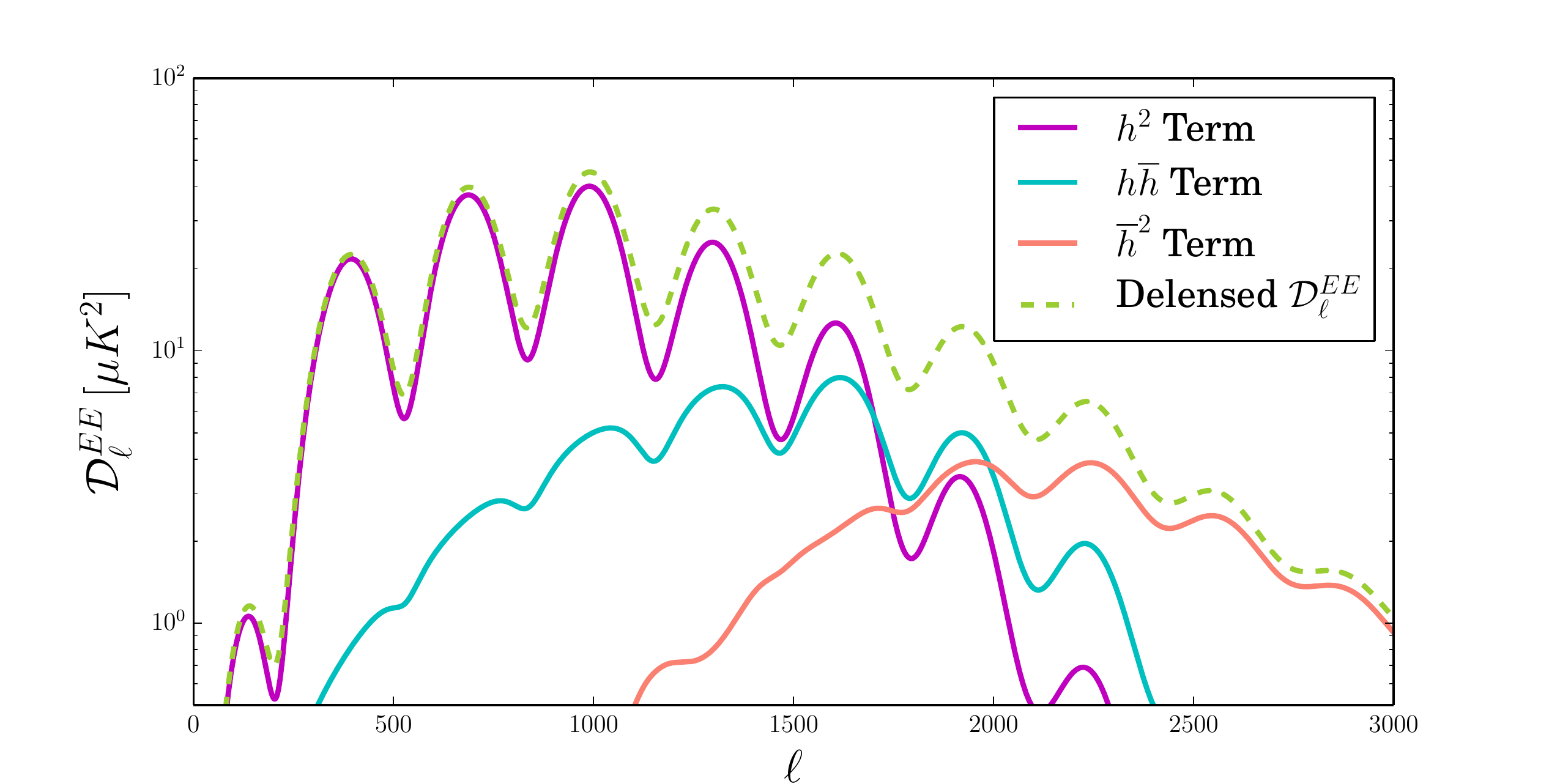}
\caption{Contributions to delensed spectrum for a Stage II experiment, where the signal-dominated modes that we delens are isolated using the $h$ filter and the noise-dominated modes we do not delens are isolated using the $\bar h$ filter.  In the power spectrum, the resulting $\bar h^2$, $\bar h h$, and $h^2$ terms  (Eq.~\ref{eqn:finalEd}) are shown in salmon, cyan, and magenta respectively.  While it is intuitively clear that the first and last terms dominate at high and low $\ell$, we see that the cross-term, $\bar h h$, can be the dominant contribution at intermediate scales.  This feature is relevant for the covariance matrices.}
\label{fig:delensing_parts_EE}
\end{center}
\end{figure}

A more detailed understanding of the spectra can be gained from isolating the various contributions.  At the power spectrum level, if we had a perfect lensing measurement, we would qualitatively expect a transition from mostly unlensed at large angular scales to mostly lensed spectra at small angular scales.  These limits are represented by the regions of the spectra where the filters take the values $h_\ell =1$ and $\bar h_\ell =1$ respectively.  Figure~\ref{fig:delensing_parts_EE} shows the contributions from each filter and indeed low $\ell$ is dominated by $h_\ell^2$ and high $\ell$ is dominated by $\bar h_\ell^2$.  However, at intermediate values the $\bar h_\ell h_\ell$ cross-term dominates, yet it has no simple interpretation as the lensed or unlensed spectrum.  This is a direct consequence of working at the map level, but this is crucial for removing the effect of the realization of $\phi$, rather than just the lensing power spectrum.  This will lead to some unexpected features in the covariance, but we will find it is necessary for improving constraints through delensing.

Our ability to delens is also affected by the noise in our lensing reconstruction.  The smearing of the acoustic peaks of the lensed spectra is due mostly to the peak of the lensing power spectrum around $L \sim 40$. At high $\ell$, there is an excess of power in the lensed spectrum as compared to the unlensed spectrum due to the presence of small scale lenses.  One can see, for example, in the $TT$ panel of Figure~\ref{fig:delensed_spectra_square} that around $\ell \sim 2500$ for the Stage IV experiment, the peak smearing has mostly been removed from the delensed spectrum, but there remains an excess of power compared to the unlensed spectrum since the small scale lenses are not resolved with high significance (see the right panel of Figure~\ref{fig:noise_power_combined}).

\subsection{Power Spectrum Covariance}


\begin{figure}[t!]
\begin{center}
\centerline{\includegraphics[width=0.8\textwidth]{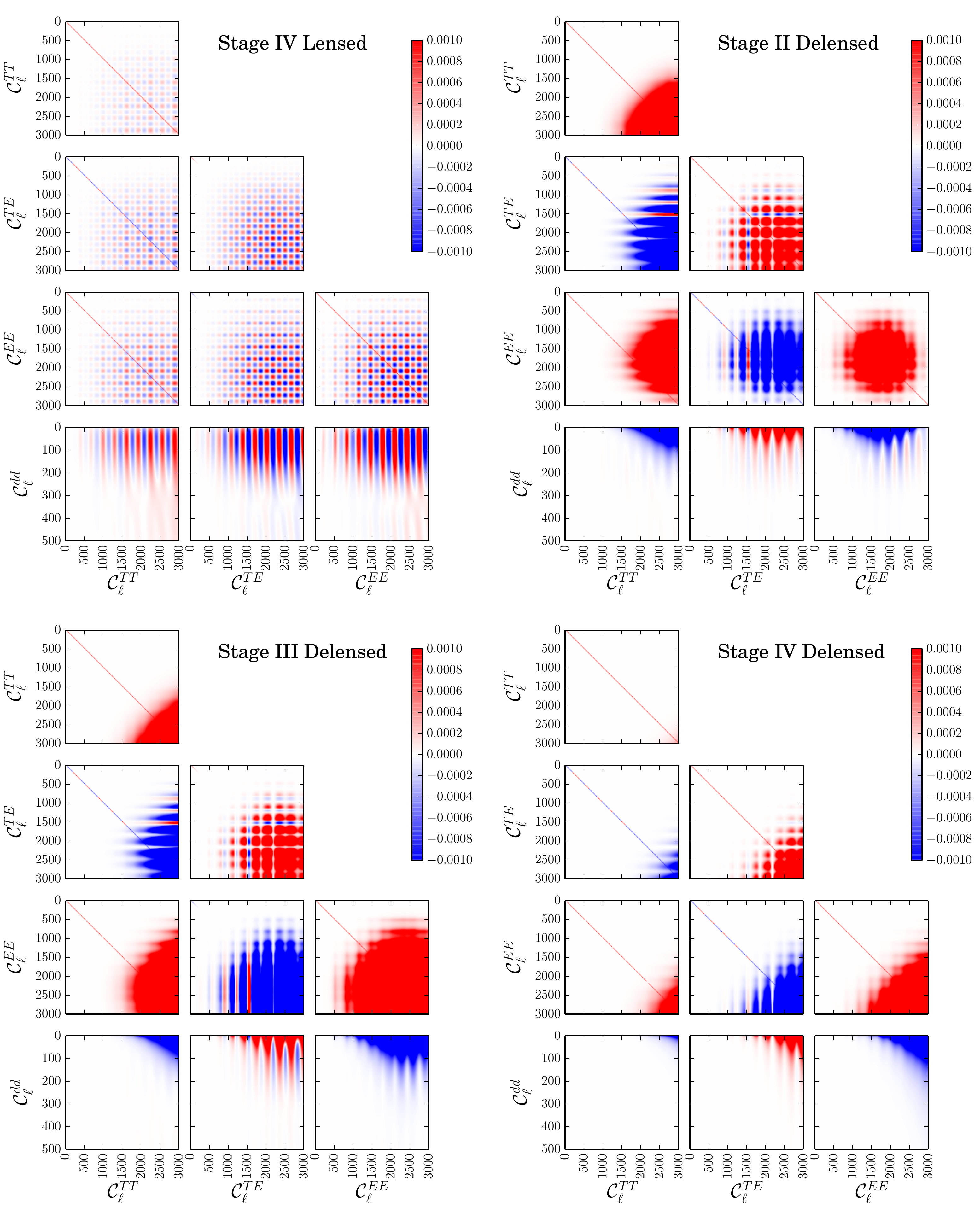}}
\caption{Correlation matrices, defined in Eq.~(\ref{eqn:corr_def}) where $d \equiv L \phi$ is the magnitude of the deflection angle. The top left panel shows the correlations for the lensed spectra with Stage-IV noise.  The other panels show the covariances for the delensed spectra for three experimental noise levels.  In the delensed covariances there is additional off-diagonal correlation at moderate values of $\ell$ beyond what appears in the lensed spectra.  One can check that this is consequence of the $\bar h h$ terms in Figure~\ref{fig:delensing_parts_EE}.  This term is only important when the noise is comparable to the signal and moves to smaller scales with higher-sensitivity data.   }
\label{fig:covs_combined_10microK}
\end{center}
\end{figure}

In order to obtain forecasts on cosmological parameter constraints, we also account for the covariances between observed CMB and lensing power spectra, calculating the  reduction in these covariances when delensing is performed.
Ignoring the non-trivial off-diagonal lens-induced covariance in a cosmological analysis will double count information encoded in the lensing field, and can overestimate the constraints on cosmological parameters that are sensitive to $C^{\phi \phi}_L$.  We will see that delensing removes these off-diagonal covariances when the noise is small; however, since our lensing map is not perfect the off-diagonal covariances are not removed perfectly, or at all for the noisiest modes.  

To compute the power spectrum covariance matrices, we will use the analytic approximation proposed by~\cite{BenoitLevy:2012va},
\bea\label{eqn:cov}
\mathrm{Cov}( C_\ell^{{\rm d}, XY}, \,  C_{\ell'}^{{\rm d},WZ})
&=& \frac{f_{\rm sky}}{2 \ell+1} \left[ C_{\ell}^{{\rm d},XW} C_{\ell}^{{\rm d},YZ} +  C_{\ell}^{{\rm d},XZ} C_{\ell}^{{\rm d},YW} \right] \delta_{\ell \,\ell'} \\
&&+ f_{\rm sky}\sum_\ml \Bigg[ \frac{\partial C_{\ell}^{{\rm d},XY}}{\partial C_\ml^{\phi \phi}} {\rm Cov}_{\ml \, \ml'}^{\phi \phi, \phi \phi} \frac{\partial C_{\ell'}^{{\rm d},WZ}}{\partial C_{\ml'}^{\phi \phi}}\Bigg] \nonumber \ , 
\eea
where $XY, WZ \in \{ TT, TE, EE \}$, $C^{{\rm d},XY}_\ell$ are the delensed spectra, and $f_{\rm sky}$ is the observed sky fraction.  In principle, our all-orders approach can also be applied directly to the covariance matrix but   is beyond the scope of this work.  The advantage of this approximate form is that it can be computed from the delensed spectra and derivatives thereof.

In practice, we will typically consider the case where we compute cross correlations of subsets of data that experience different realizations of the noise.  This removes the noise when computing these cross-spectra.  For the covariance matrix, this amounts to considering $X=W= T$ and $Y=Z= T'$ where $C^{TT'}_\ell$ is the delensed spectrum with $C^{{\rm N}, TT'}_\ell = 0$.  In Equation~(\ref{eqn:cov}), we see that the noise still enters in the diagonal term via $C^{XW = TT}_\ell C^{YZ=T'T'}_\ell$ but does not enter in the off-diagonal terms.  

Calculating the derivative of the CMB spectra with respect to the lensing spectrum is tedious but straightforward.  Essentially, it is a simple application of the result for the lensed $TT$ spectrum
\bea
\frac{\partial \tilde C_{\ell}^{TT}}{\partial C^{\phi \phi}_\ml }& =&  \int d^2 r \frac{d^2 \ell_1 }{(2\pi)^2}   C_{\ell_1} e^{-i (\vl - \vl_1) \cdot \r} \frac{\partial }{\partial C^{\phi \phi}_\ml }  \left[  e^{-\frac{\ell^2_1}{2} (C_0(0)-C_0(r)+ \cos 2\varphi_1 C_2(r))}  \right] \\
&=& \int d^2 r \frac{d^2 \ell_1 }{(2\pi)^2}   C_{\ell_1} \frac{\ell^2_1 \ml^3}{2}\bigg(J_0(\ml r) -1  - \cos 2\varphi_1 J_2(\ml r) \bigg) e^{-i (\vl - \vl_1) \cdot \r} e^{-\frac{\ell^2_1}{2} (C_0(0)-C_0(r)+ \cos 2\varphi_1 C_2(r))} \nonumber  \ .
\eea

The full result for the delensed $TT$ spectrum is shown in Appendix~\ref{app:covmat} and is trivially generalized to the polarization spectra.  As with the power spectra, we expand to first order in $C_2(r)$ for the numerical computation of the covariance matrices.

The covariance is most easily visualized through the dimensionless correlation matrix shown in Figure~\ref{fig:covs_combined_10microK},  defined as 
\beq
\label{eqn:corr_def}
\mathrm{Corr}\left(C_\ell^{XY}, \,  C_{\ell'}^{WZ}\right) \equiv \frac{\mathrm{Cov}( C_\ell^{XY}, \,  C_{\ell'}^{WZ}) }{ \sqrt{\mathrm{Cov}( C_\ell^{XY}, \,  C_{\ell}^{XY}) \,  \mathrm{Cov}( C_{\ell'}^{WZ}, \,  C_{\ell'}^{WZ}) } } \, , 
\eeq
where the contribution from the CMB noise is included in the covariance.  For the lensed spectra, we see the characteristic checkerboard pattern in the $TT$, $TE$, and $EE$ spectra from the lens-induced mode coupling~\cite{BenoitLevy:2012va}.  We also see that the correlation between lensing power spectra and the CMB power spectra is dominated by the low-$L$ lensing modes over a wide range of scales in the CMB~\cite{Schmittfull:2013uea}.  In contrast, the delensed spectra show essentially none of these features at low $\ell$ where we expect the delensing to be effective.  This is precisely what we would expect by removing the effect of lensing.  However, at moderate values of $\ell$ we see that delensing has produced large off-diagonal correlations that have no analog in the lensed spectra.

The large off-diagonal terms are a result of the $h_\ell \bar h_\ell$ term of Eq.~\ref{eqn:ttresult} shown in Figure~\ref{fig:delensing_parts_EE}.  Effects of lensing that remain in the $\bar h$-filtered maps lead to a broadband lensing contribution in the spectra when cross correlated with the $h$-filtered maps.  This in turn causes off-diagonal covariance between modes which contain a significant contribution from the $h_\ell \bar h_\ell$ term.  This explains why these off-diagonal correlations are confined to intermediate value of $\ell$, as it is due to the transition from $h_\ell$ to $\bar h_\ell$ where both are important.  Furthermore, these regions would not appear if we were simply to set $\bar h_\ell = 0$ throughout.  However, our goal is to get the best possible measurement of cosmological parameters, not minimize the off-diagonal correlations.  We will see in Section~\ref{sec:params} that these regions do not negatively impact the forecasts and, in fact, attempts to remove the off-diagonal correlations by taking $\bar h_\ell = 0$ significantly weaken the constraints.  As a result, while these off-diagonal features may seem undesirable, they are an inevitable consequence of choices that optimize delensing both for map making and for cosmological parameters.

\section{Delensing and Cosmological Parameters}
\label{sec:params}

The primary applications of our all-orders delensed spectra are to forecasting and data analysis.  A real experiment will produce temperature, polarization, and lensing maps with the goal of measuring cosmological parameters.  We have seen that delensing does sharpen acoustic peaks and removes some of the lensing-induced non-Gaussian covariance.  Of course, the practical value of delensing is seen in the error bars of cosmological parameters.  In this section, we will explore forecasts using delensed spectra for a variety of cosmological parameters.  We will see that forecasts always improve with delensed spectra, providing a more unambiguous reason to delens the CMB maps (see also Appendix~\ref{app:info} for discussion).

\subsection{Forecasting Methodology}\label{sec:forecastmethod}

All unlensed, lensed, and deflection power spectra used in forecasts are computed using CAMB \cite{Lewis2000}.  Delensed spectra are computed from the CAMB output by using Eqs.~(\ref{eqn:delta_xiTd}), (\ref{eqn:delensed_temp_spec}), and (\ref{eqn:delta_xiplusd}-\ref{eqn:delensed_pol_spec}).  

For the noise in the CMB survey, we assume Gaussian noise spectra of the form 
\begin{equation}\label{eqn:noise_power_def}
	C_\ell^{TT,\rm{N}} = \Delta_T^2 \exp \left(\ell(\ell+1)\frac{\theta_{\mathrm{FWHM}}^2}{8\log 2}\right) \, , 
\end{equation}
where $\Delta_T$ is the instrumental noise in $\mu$K-radians and $\theta_{\mathrm{FWHM}}$ is the beamsize in radians. 
We assume fully polarized detectors, such that the polarization noise spectra are $C_\ell^{EE,\rm{N}} = C_\ell^{BB,\rm{N}} = 2  C_\ell^{TT,\rm{N}}$.  

For the noise in the lensing reconstruction, we assume that the given CMB survey is used to obtain a lensing map with standard quadratic estimator techniques \cite{Hu:2001kj}.  We note, however, that maps obtained from tracers of large-scale structure~\cite{Smith2010}, such as the emission from the cosmic far-infrared background~\cite{Sherwin:2015baa}, can yield higher-fidelity maps of lensing than those obtained internally from the CMB for some upcoming experiments.   We use the minimum variance quadratic estimator, which combines information in the lensed temperature and polarization fields.  For the $EB$ estimator, which dominates the lensing information in a high-sensitivity experiment, our calculation of the lensing reconstruction noise includes the improvement from iterative delensing~\cite{Smith2010}.  We use this iterative technique only for the purpose of minimizing lensing reconstruction noise, and we use our all-orders method when computing delensed spectra. 

Although we model the lens reconstruction noise as coming from the given CMB survey, we are ignoring terms that would arise in both the spectra and covariances if the lensing field were obtained from quadratic combinations of the same $T$ and $E$ modes that are being delensed.  Including these terms will generically lead to biases on the delensed power spectra.  These biases can be avoided using independent maps of the lensing field such as those from large scale structure~\cite{Smith2010,Sherwin:2015baa} or by using CMB modes in the lens reconstruction that are disjoint from those being delensed.  This latter technique is  analogous to a  method for avoiding the  bias on measured lensing  power auto-spectra originating from the  disconnected CMB four-point function~\cite{sherwin10, vanEngelen:2012va}.  Avoiding these biases also has the benefit of avoiding off-diagonal lens-lens covariance~\cite{Hanson:2010rp} as well as obviating the need for additional terms in the temperature- and polarization-lens power cross-covariance~\cite{Schmittfull:2013uea} in Eq.~(\ref{eqn:cov}).

In order to forecast constraints on a set of cosmological parameters ${\lambda_i}$, we compute Fisher matrices using
\begin{equation}\label{eqn:fisher_matrix}
F_{ij} = \sum_{\ell,\ell'} \sum_{WX,YZ} {\partial C_\ell^{XY} \over \partial \lambda_i} \mathrm{Cov}^{-1}( C_\ell^{XY}, C_{\ell'}^{WZ}) {\partial C_{\ell'}^{WZ} \over \partial \lambda_j}.
\end{equation}
In this sum over power spectra, as well as when computing reconstructed lensing maps, we take $\ell_\mathrm{min} = 30$ and $\ell_\mathrm{max}=5000$, except for $TT$ spectra, for which we use $\ell_\mathrm{max}^{TT}=3000$, due to the presence of foregrounds in the CMB temperature, such as  radio and emission high-redshift galaxies and the thermal and kinetic  Sunyaev-Zel'dovich effects. We take the sky fraction to be $f_\mathrm{sky} = 0.7$, and we assume a 1 arcminute beam.

We have simplified the calculation of the Fisher matrix in Equation (\ref{eqn:fisher_matrix}) by ignoring derivatives of the inverse covariance.  This approximation is typically very reliable, but turns out to create complications with $C^{{\rm d}, TE}_\ell$.  Specifically, when $h_\ell \neq \hP_\ell$ delensing changes the RMS power in $C^{TE}_\ell$ by an amount that depends on $C_0(0)$, unlike $C^{TT}_\ell$ and $C^{EE}_\ell$ which conserve total power.  This can be seen as excess power at high-$\ell$ in Figure~\ref{fig:delensed_spectra_square}.  In forecasts, this excess power should cancel between signal and the noise but this cancellation is missed in the Fisher matrix when dropping derivatives of the covariance.  To avoid this technical challenge we will use $h^T_\ell = \hP_\ell$ in $C^{{\rm d}, TE}_\ell$.  This choice has the additional advantage that the  $T$ and $E$ fluctuations are shifted by the same amount in the delensing process, preserving the $TE$ correlation.  In principle, one might imagine improvements in parameter constraints by using $h^T_\ell$ instead but we will see that at Stage~IV noise levels there is little room for improvement in most parameters.

Fiducial cosmological parameters and step sizes for numerical derivatives are listed in Table~\ref{table:cosmo_fiducial}.  We use $TT$, $TE$, $EE$, and $dd$ power spectra in all forecasts ($d \equiv \ell \phi$).  For the covariance matrix $\mathrm{Cov}( C_\ell^{XY}, C_{\ell'}^{WZ})$ for lensed and delensed CMB spectra, we include non-Gaussian covariance term given in Eq.~(\ref{eqn:cov}).

Note that we have not included external data such as BAO information or a prior on the optical depth $\tau$ in order to highlight the various aspects of delensing.  On the other hand, this choice results in forecasts which are weaker than other published results, especially for $\sum m_\nu$.

\begin{table}[t!]
\begin{center}
 \begin{tabular}{ll cc} 
 \toprule
   Parameter & Symbol    &   Fiducial Value      & Step Size     \\ [0.5ex] 
 \midrule
Physical cold dark matter density &   $\Omega_c h^2$ &   0.1197 	            & 0.0030 	    \\ 
Physical baryon density &   $\Omega_b h^2$ &   0.0222 	            & $8.0\times10^{-4}$ 	    \\
Angle subtended by acoustic scale &   $\theta_s$     &   0.010409 	            & $5.0\times10^{-5}$ 	    \\
Thomson optical depth to recombination &   $\tau$         &   0.060 	            & 0.020 	    \\
Primordial scalar fluctuation amplitude &   $A_s$          &   $2.196\times10^{-9}$  & $0.1\times10^{-9}$ 	    \\
Primordial scalar fluctuation slope &   $n_s$          &   0.9655 	            & 0.010 	    \\
Sum of neutrino masses &   $\sum m_\nu$ [eV] &   0.060 	            & 0.020 	    \\
Number of radiation-like species &   $\Nf$          &   3.046 	            & 0.080 	    \\
\hline
Primordial helium fraction &   $Y_p$          &   0.2467 	            & 0.0048 	    \\
  \bottomrule
\end{tabular}
\caption{Fiducial cosmological parameters and step sizes for numerical derivatives used in forecasts.  We consider the 8-parameter $\Lambda$CDM+$\sum m_\nu$+$\Nf$ model with $Y_p$ set by BBN consistency for the given values of $\Omega_b h^2$ and $\Nf$, as well as the 9-parameter $\Lambda$CDM+$\sum m_\nu$+$\Nf$+$Y_p$ model in which $Y_p$ is free.  
}
\label{table:cosmo_fiducial}
\end{center}
\end{table}

\subsection{Implications}

\begin{figure}[t!]
\begin{center}
\includegraphics[width=0.85\textwidth]{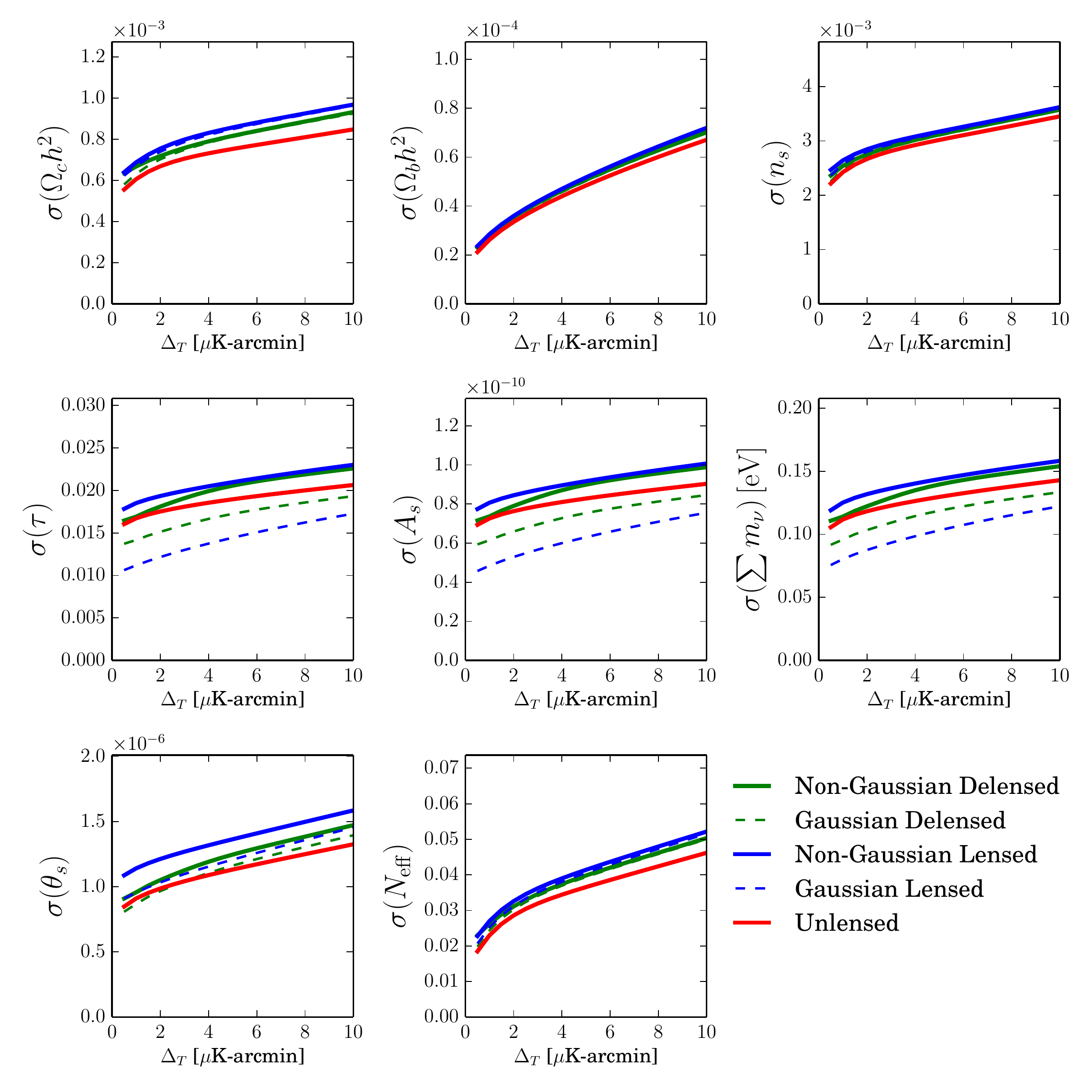}
\caption{Forecasts showing the improvement from delensing for the 8-parameter $\Lambda$CDM+$\sum m_\nu$+$\Nf$ model with $Y_p$ fixed by BBN consistency.  One can see from the middle row that when neglecting lensing-induced covariance, errors on parameters which depend on lensing information are underestimated and delensing appears to worsen constraints.  When non-Gaussian covariances are properly included, delensing improves constraints for all parameters and all noise levels.  At low noise levels, we see that forecasts saturate the unlensed forecasts for several parameters.}
\label{fig:BBN_param_constraints}
\end{center}
\end{figure}

The behavior of cosmological parameters in our forecasts ultimately splits into two categories: parameters that are constrained by the primary CMB (e.g.~$\theta_s$, $\Neff$, and $Y_p$) and those that benefit from lensing information (e.g.~$\tau$, $A_s$, and $\sum m_\nu$).  These extreme cases will both be important in highlighting how delensing affects information and the role played by the non-Gaussian covariances.  We find that delensing always increases the Fisher information.  We expand on some of these points in Appendix~\ref{app:info}.

The qualitative effects of delensing are seen most easily in Figure~\ref{fig:BBN_param_constraints} which shows forecasted errors on an 8-parameter $\Lambda$CDM+$\sum m_\nu$+$\Nf$ model, where $Y_p$ is fixed to be consistent with the predictions of standard big bang nucleosynthesis (BBN) for the given values of $\Omega_bh^2$ and $\Nf$.  Constraints on all the parameters are seen to improve at least marginally with delensing, in many cases coming close to the unlensed constraints in the limit of no instrumental noise.  This confirms for this model that delensing always increases Fisher information.  However, we also see that the improvements for a given parameter depend sensitively on the detailed effect it has on the power spectrum.  One can typically tell from the unlensed and lensed constraints which parameters ought to improve with delensing, although $\Omega_c h^2$ shows that this is by no means guaranteed.  

\begin{figure}[t!]
\begin{center}
\includegraphics[width=\textwidth]{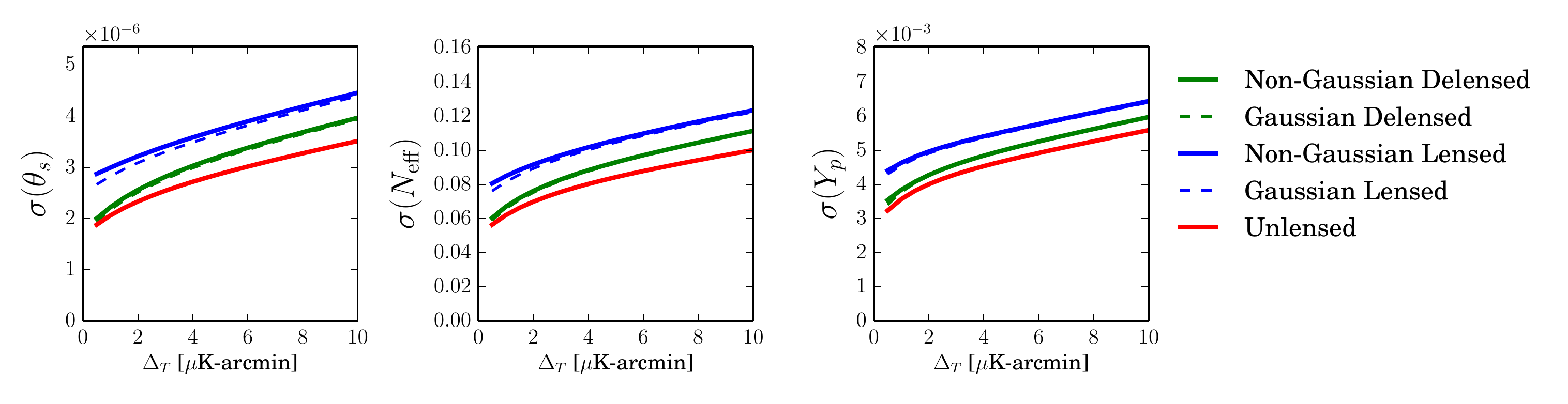}
\caption{Forecasts for a subset of parameters which benefit from peak sharpening in the 9-parameter $\Lambda$CDM+$\sum m_\nu$+$\Nf$+$Y_p$ model.  In this regime, for the parameters shown, the benefit of delensing is roughly equivalent to a two-fold improvement in sensitivity if no delensing were performed.}
\label{fig:delensing_improvement}
\end{center}
\end{figure}

Delensing most clearly improves the measurement of parameters that affect the acoustic oscillations, as was anticipated in Sec.~\ref{sec:intro}.  This is seen  explicitly in Figure~\ref{fig:delensing_improvement}, which shows the forecasts for $\theta_s$, $\Neff$, and $Y_p$ for the 9-parameter model where $Y_p$ is free.  The angular scale of the acoustic horizon, $\theta_s$, directly determines the peak locations and benefits most from delensing, due to peak sharpening.  This is clearly seen in the forecasts, as the measurement of $\theta_s$ smoothly interpolates between the lensed and unlensed forecasts as we lower the noise of the experiment (and therefore the lensing reconstruction noise as seen in Figure~\ref{fig:noise_power_combined}).  In this case, delensing is literally playing the same role as BAO-reconstruction in sharpening the BAO peak of the correlation function and reducing the error in $\theta_s$, as shown in Figure~\ref{fig:BAO_analogy}.

Similar behavior is seen for other cosmological parameters that affect the acoustic peaks, but the effect is most significant when we isolate the effect on the acoustic oscillations.  A parameter of particular interest is $\Neff$ which affects the peak locations~\cite{Bashinsky:2003tk,Follin:2015hya,Baumann:2015rya} but also alters the damping scale~\cite{Bashinsky:2003tk,Hou:2011ec}.  The effect on the damping tail is degenerate with $Y_p$ and by marginalizing over $Y_p$ we can isolate the phase shift.   In particular we see that the error on $\Nf$ in the 9-parameter $\Lambda$CDM+$\sum m_\nu$+$\Nf$+$Y_p$ model for a Stage IV experiment improves from $\sigma(\Nf) = 0.085$ with lensed spectra to $\sigma(\Nf) = 0.067$ with delensed spectra, an improvement of 21\%. Figure~\ref{fig:delensing_improvement} also shows the constraints on $\Neff$ and $Y_p$, showing the  same improvement for these parameters that we observed with $\theta_s$.  We also see that the residual non-Gaussian covariances induced by our filtering scheme shown in Figure~\ref{fig:covs_combined_10microK} have no meaningful impact on our constraints on these parameters.  Specifically, the forecasts with Gaussian covariances give essentially the same results indicating a negligible effect form the off-diagonal terms. Non-Gaussian covariances will be important for parameters that are sensitive to the lensing power spectrum, as we discuss below and in Appendix~\ref{app:info}.

\begin{figure}[t!]
\begin{center}
\includegraphics[width=0.6\textwidth]{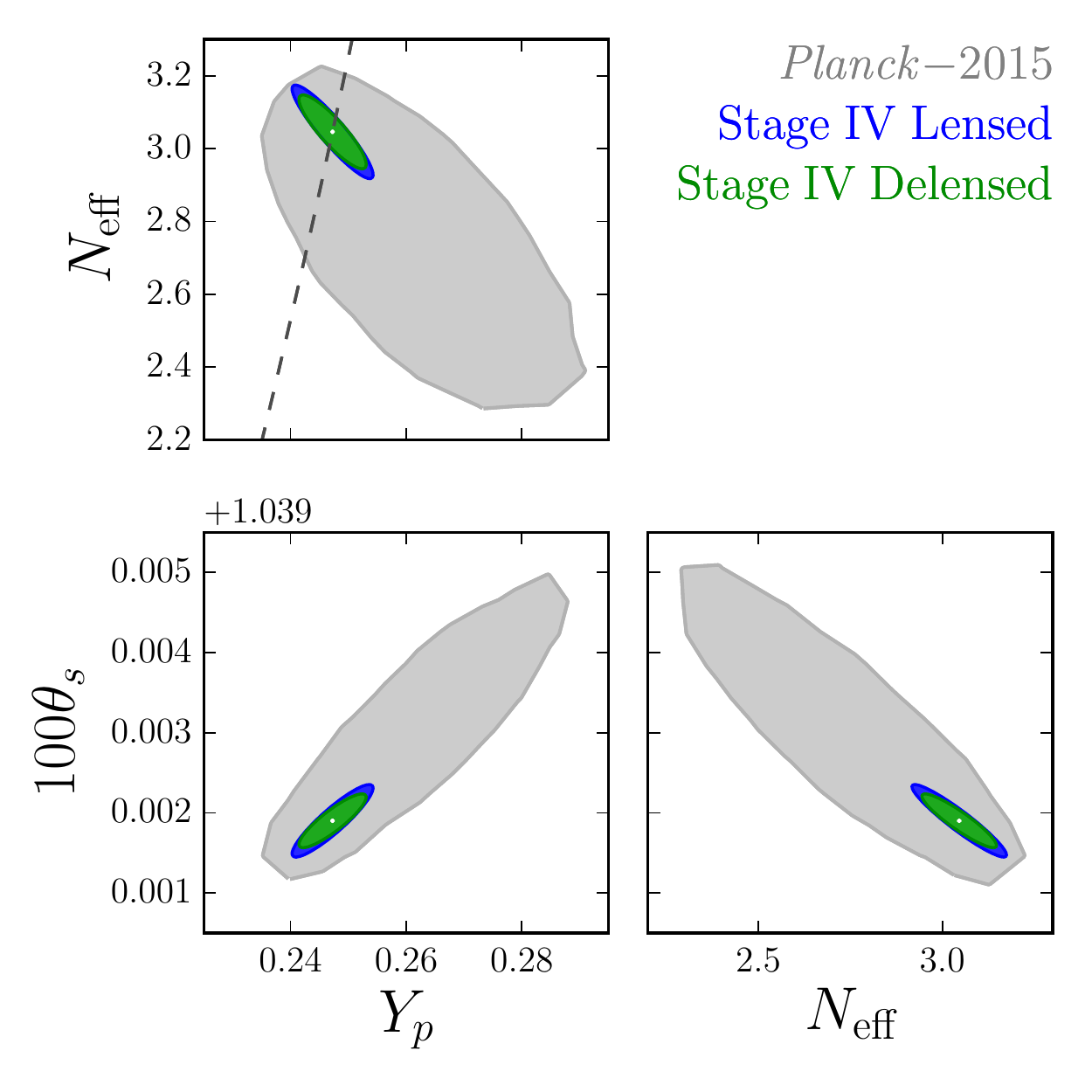}
\caption{Forecasts showing the improvement from delensing for 2d ellipses of 68\% confidence levels  in the 9-dimensional $\Lambda$CDM+$\sum m_\nu$+$\Nf$+$Y_p$ parameter space for a Stage IV experiment, as well as current constraints from {\it Planck}~\cite{planck15lensing}.  Specifically, we use the ``base nnu yhe plikHM TTTEEE lowTEB'' chains provided by the Planck team at \url{http://wiki.cosmos.esa.int/planckpla2015/index.php/Cosmological_Parameters}. 
  In the $\Neff$--$Y_p$ panel we see that delensing makes a more significant impact along the direction of the degeneracy.  This is consistent with the expectation that the measurement of the phase shift is important for breaking the degeneracy and this measurement improves with peak sharpening.  The dashed line shows the prediction assuming BBN consistency, which gives $Y_p$ for a given value of $\Neff$ (assuming $\Neff$ does not vary in time).  This line is not aligned with the degenerate direction which explains why there is little improvement in the constraint on $\Neff$ from delensing when assuming BBN consistency, as shown in Figure~\ref{fig:BBN_param_constraints}.  }
\label{fig:Neff_Yp}
\end{center}
\end{figure}

We can further isolate the effect of delensing on the phase shift by examining the contours in the $\Neff$-$Y_p$ plane.   The phase shift induced by $\Neff$ breaks the degeneracy between $\Neff$ and $Y_p$ and therefore we expect delensing to have a larger effect along the line of degeneracy.  Figure~\ref{fig:Neff_Yp} shows that this is precisely what happens in our forecasts.  We also see that prediction of BBN consistency, in which ~$Y_p$ is determined in terms of $\Neff$, assuming otherwise standard BBN, is not aligned with to the degenerate direction.  As a consequence, forecasts for $\Neff$ that assume BBN consistency show only a marginal improvement in going from lensed to delensed spectra (see Figure~\ref{fig:BBN_param_constraints}).  A larger difference between lensed and unlensed forecasts for $\Neff$ had been noticed previously~\cite{Baumann:2015rya} although in that case it was likely due to information in the damping tail of the unlensed spectra whereas the delensed spectra discussed here are much closer to the lensed spectra at small angular scales\footnote{It is also worth noting that unlike in Ref.~\cite{Baumann:2015rya} we have set $\ell_{\rm max}^{TT} =3000$ which removes all of the high-$\ell$ information in temperature.}.

For parameters that are directly influenced by $C^{\phi \phi}_{L}$, the benefit of delensing is less clear.  Delensing {\it removes} the information about the lensing potential and one could worry that delensing could weaken constraints.  The intuitive reason that delensing does not weaken constraints is that $\phi$ is a Gaussian field and therefore all cosmological information is encoded in $C^{\phi \phi}_\ell$, which we also include in our likelihood.  Any information that we are removing from the CMB spectra is then being included through the lensing power spectrum.  We see this explicitly in the second row of Figure~\ref{fig:BBN_param_constraints}, where the constraints on $\tau$, $A_s$, and $\sum m_\nu$ are all seen to improve with delensing and even saturate the unlensed forecasts.  However, in this case we see that it is always important to include non-Gaussian covariances.  Failing to include lensing-induced non-Gaussian covariances gives overly optimistic constraints and it would appear that delensing actually weakens parameter constraints.  The unlensed spectra do not contain any non-Gaussian covariances and it is therefore noteworthy that the delensed constraints reproduce the unlensed result at low noise.  This shows that the residual off-diagonal covariances in Figure~\ref{fig:covs_combined_10microK} have no meaningful impact on these measurements.

\begin{figure}[t!]
\begin{center}
\includegraphics[width=0.85\textwidth]{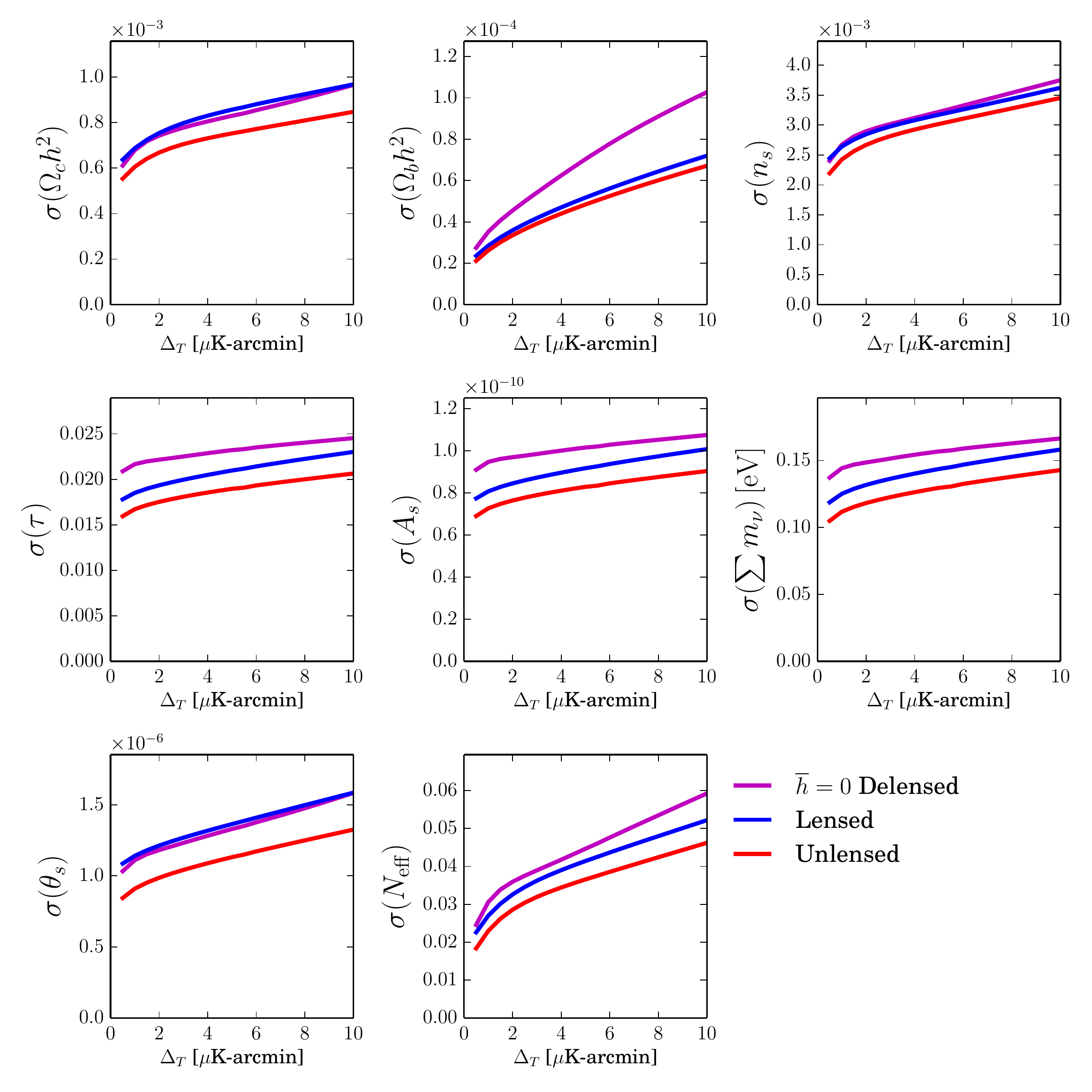}
\caption{Forecasts demonstrating the drawbacks of a filtering scheme in which only the imaged CMB modes are delensed, i.e., setting $\bar h = \bhP = 0$ while $h$ and $\hP$ are given by Eqs.~(\ref{eqn:optimal_temp_filter}) and (\ref{eqn:optimal_pol_filter}) respectively.  One can clearly see that with this choice of filtering, parameter constraints with delensing are worse than when using the filtering scheme that conserves total power, and in fact worse than performing no delensing for most parameters.}
\label{fig:hbar_zero_param_constraints}
\end{center}
\end{figure}

\vskip 10pt
Our forecasts clearly demonstrate the benefits of delensing.  However, one might question whether the full filtering scheme introduced in Equation~(\ref{eqn:Tddefinition}) is necessary to achieve similar results.  Of particular concern is the need for both the $\bar h$ and $h$ terms in our map.  In Figure~\ref{fig:covs_combined_10microK}, we saw that the interplay between these two terms added significant off-diagonal correlations that are not present in the unlensed or lensed covariances.  One could eliminate much of this by setting $\bar h_\ell = 0$ but leaving $h_\ell$ unchanged (this does not conserve total power because $h_\ell \neq 1$).  The downside of this procedure is that we are suppressing both the signal and the noise and one may ultimately be losing information.  Forecasts shown in Figure~\ref{fig:hbar_zero_param_constraints} demostrate that this is indeed what occurs and setting $\bar h =0$ significantly reduces the sensitivity of the experiment.  In this sense, we see that our filtering scheme is important for achieving the benefits of delensing.

\section{Conclusion}
\label{sec:concl}

In this paper, we have shown that future CMB experiments will be  sufficiently sensitive to CMB lensing that the delensing of all of the spectra (and not just $B$ modes) can meaningfully improve the constraints on cosmological parameters.  Delensing sharpens in the acoustic peaks, improving the measurement of peak locations and any cosmological parameters that affect the acoustic structure.  Delensing also removes the lens-induced covariances for the modes measured with high significance.

We have shown how to compute the predictions for the delensed spectra and covariances to all orders in the lensing potential.  We used these results to model the impact of delensing on cosmological parameters of interest.  The most notable improvements occurred for parameters sensitive to peak locations and associated parameters that would otherwise be degenerate.  In $\Lambda$CDM, the most dramatic improvements occurred for $\theta_s$ which is directly a measurement of the peak locations.  When $Y_p$ and $\Neff$ are both free, the phase shift due to $\Neff$ breaks the degeneracy between these two parameters and delensing is seen to substantially improve error bars, showing an improvement of roughly 20\% for both parameters with a Stage IV experiment when compared to forecasts with lensed spectra.  More generally, we show that when the residual lens-induced covariances are included, Fisher information always increases when using delensed, rather than lensed, spectra. 

Looking forward, delensed spectra will ultimately be necessary not just for forecasting but also for any likelihood analysis with delensed data.  However, unlike lensed or unlensed spectra, the theoretical predictions depend also on the experimental noise.  The analysis presented here computes these spectra for more idealized experiments.  In principle, the approach taken here will generalize to any experiment, but real data  may violate some of the technical assumptions needed to simplify our analytic predictions.  More optimistically, we did not fully solve the problem of how to optimize our filters to maximize the Fisher information and one might imagine even more information may yet be available.  As delensing of the CMB becomes more commonplace, these and other extensions of this work will deserve further exploration.

\vskip23pt
\paragraph{Acknowledgements}
We thank Daniel Baumann, Raphael Flauger, Marcel Schmittfull, Neelima Sehgal,  Blake Sherwin, and Kendrick Smith for helpful discussions. D.G.~and J.M.~also thank Daniel Baumann and Benjamin Wallisch for collaboration on related work that inspired this project.  D.G.~was supported by an NSERC Discovery Grant and the Canadian Institute for Advanced Research. J.M.~was supported by the Vincent and Beatrice Tremaine Fellowship.

\newpage
\appendix

\section{Gradient Expansion} \label{app:grad}

In Section~\ref{sec:theory}, we presented two procedures for delensing.  In the limit of a noiseless measurement, these two procedures are
\beq
\Td_{1}{}^{, (J)} (\x) = \int d^2 x' J(\x') \delta(\x-\x'-\va(\x')) \tilde T(\x') \qquad {\rm and } \qquad \Td_2(\x) = \tilde T(\x-\va(\x)) \ ,
\eeq
where $J(\x) = \det \partial_i (x_j + \alpha_j(\x))$.  The advantage of working with $\Td_{1}{}^{,(J)}$ is that it reproduces the unlensed map, while $\Td_2$ disagrees with the unlensed map due to gradients.  In practice, $\Td_{1}{}^{,(J)}$ is a difficult procedure to implement, in part because we must compute the determinant $J(\x)$ for the observed map $\vao(\x)$ (see Appendix~\ref{app:exact} for details).  Yet, we are dropping ``gradients" so one might imagine that we can approximate $\Td_{1}{}^{ (J)}$ by taking $J(\x) \to 1$.  In this appendix, we will show that error made by dropping gradients in $\Td_2$ is acceptably small while $\Td_1{}^{,(J=1)}$ is not sufficiently accurate.

First let us estimate the error made by setting $J(\x) = 1$ by defining
\beq
\Td_1(\x) = \int d^2 x'  \delta(\x-\x'-\va(\x')) \tilde T(\x')  \ ,
\eeq
which is given in terms of harmonics by
\beq
\Td_{\vl, 1} = \int d^2 x' \frac{d^2 \ell_1}{(2\pi)^2}  e^{-i (\vl-\vl_1) \cdot (\xp + \va(\x'))} T_{\vl_1} \ .
\eeq
The residual lensing in the power spectrum due to the gradients (i.e the error in setting $J \to 1$) is given by
\beq
\Delta C_{1,\ell}^{\rm d} = \int d^2 r \frac{d^2 \ell'}{(2\pi)^2} C_{|\vl - \vlp|}    e^{-i \vlp \cdot \vec r} \Big[e^{- \frac{\ell'{}^2}{2} (C_0(0)-C_0(r)+ \cos 2\varphi' C_2(r))} -1 \Big] \ , 
\eeq
where we made the change of variables $\vlp = \vl-\vl_1$ and defined $\hat{\ell}\mkern2mu\vphantom{\ell}' \cdot \hat r \equiv \cos \varphi'$ and $\Delta C_\ell^{\rm d} \equiv C_\ell^{\rm d} - C_\ell$.  To estimate the size of the error, we will Taylor expand in $\sigma(r) = C_0(0) -C_0(r)$ and drop $C_2(r) \ll C_0(r)$ to get
\bea
\Delta C_{1,\ell}^{\rm d} &\approx&  \int d^2 r \frac{d^2 \ell'}{(2\pi)^2} C_{|\vl - \vlp|}    e^{-i \vlp \cdot \vec r} \frac{\ell'{}^2}{2}\big[C_0(r)-C_0(0)\big] \nonumber \\ 
 &\approx& \int  \frac{d^2 L}{(2\pi)^2} C_{|\vl - \vL|} \frac{L{}^4}{2} C_{L}^{\phi \phi} \label{eq:Td1}
\eea
We should compare this to the perturbative correction from lensing
\bea\label{eqn:pertlens}
\Delta \tilde C_\ell &\approx& \int  \frac{d^2 L}{(2\pi)^2}  \, \frac{L{}^2}{2} C_{L}^{\phi \phi} \Big[\left|\vl-\vL\right|^2 C_{|\vl - \vL|}  - \ell^2 C_\ell  \Big] \ .
\eea
If the integrals in Eqs.~(\ref{eq:Td1}) and~(\ref{eqn:pertlens}) are both dominated in by $L  \ll \ell$, then the gradient term, Eq.~(\ref{eq:Td1}), would be suppressed by $(L/\ell)^2$.  While this is true for the effect of lensing in Eq.~(\ref{eqn:pertlens}), $L^6 C^{\phi \phi}_{L}$ grows with increasing $L$ and the gradient term is actually dominated by $L \sim \ell$ and therefore is unsuppressed.  We see that $J(\x)$ is not a small correction but is crucial\footnote{In practice, we also filter $\vao$ such that we are often not integrating up to $L \sim \ell$ for $\ell > 1000$.  Nevertheless, the integral is still dominated by the largest $L$ allowed by the filter.  In this sense, the intuition that the gradients are controlled by the peak of $C^{\phi \phi}_{L}$ around $ L \sim 40$ is not correct and the suppression only from the filter is not sufficiently accurate.} for avoiding this large error.

\begin{figure}[t!]
\begin{center}
\includegraphics[width=0.7\textwidth]{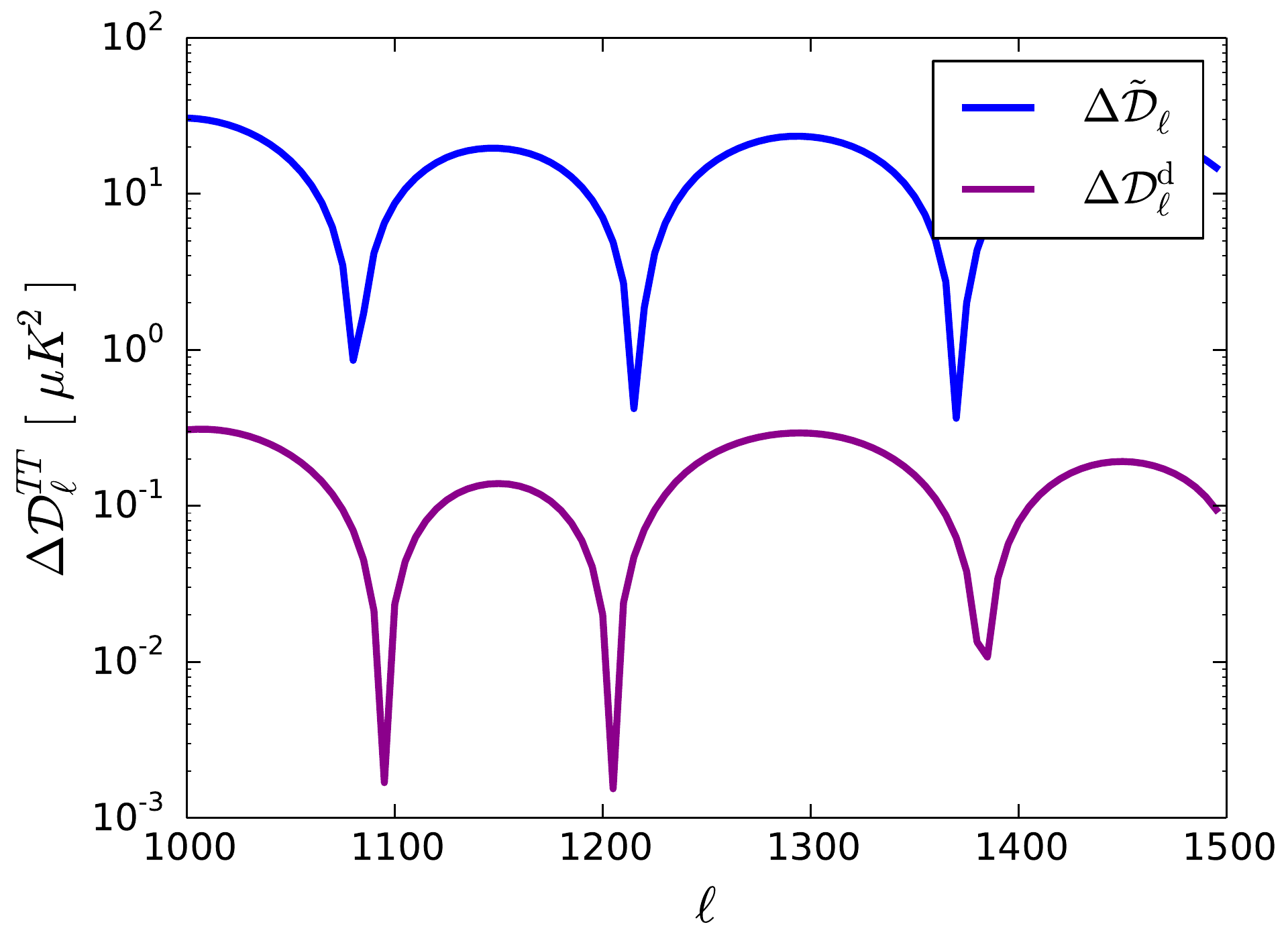}
\caption{Comparison of the error due to ignoring gradients from Equation~(\ref{eqn:pertgrad}) compared to perturbative effect of lensing from Equation~(\ref{eqn:pertlens}).  We see that gradients are roughly two orders of magnitude smaller than the correction from lensing.}
\label{fig:grad}
\end{center}
\end{figure}

Now let us compare to the error made by dropping gradients in $\Td_2$.  To first order in gradients we have
\beq
\Td_{\vl, 2} = \int d^2 x' \frac{d^2 \ell_1}{(2\pi)^2}  e^{-i (\vl-\vl_1) \cdot \xp} e^{i \ell_1^j \nabla^i \a_j(\x') \a_i(\x')} T_{\vl_1} \ .
\eeq
It is easy to check that translation and rotation invariance requires $\langle \nabla^i \a_j(\x') \a_i(\x') \rangle = 0$.  Therefore the leading correction is ${\cal O}(\alpha^4)$ and is given by 
\bea
\Delta  C^{\rm d}_{2,\ell} &\approx& \int  \frac{d^2 \ell' d^2 r}{(2\pi)^2} \left|\vl - \vlp\right|^2 C_{|\vl - \vlp|} e^{i \vlp\cdot \vec r} \frac{1}{4}  \Big[C^{\nabla^2}_0(r) C_0(r) -C^{\nabla^2}_0(0) C_0(0) \Big] \ ,
\eea
where we have dropped $C_2(r)$ and similar terms and defined
\beq
C^{\nabla^2}_0(r) = \int \frac{d^2 L}{(2\pi)^2} L^4 C_L^{\phi \phi} e^{i \vL \cdot \vec r} \ .
\eeq
We can integrate over $r$ and one of the momenta to find
\bea\label{eqn:pertgrad}
\Delta  C^{\rm d}_{2,\ell}  &\approx& \int  \frac{d^2 L_1 d^2 L_2}{(2\pi)^4} \frac{L_2^2 }{2} C_{L_2}^{\phi \phi} \,  \frac{L_1^4}{2} C_{L_1}^{\phi \phi} \, \Bigg[ \left|\vl - \vL_1-\vL_2\right|^2 C_{|\vl - \vL_1-\vL_2|} -\ell^2 C_\ell \Big] \ ,
\eea
Like $\Td_1$, we see that the contribution from the derivative causes the $L_1$ integral to be peaked at $L_1 \sim \ell$ and is therefore unsuppressed.  However,  $\Td_2$ is additionally suppressed by $C_0(0) \simeq 10^{-7}$ which keeps the effect from the gradients small compared to the effect of lensing.  Equation~(\ref{eqn:pertgrad}) can be integrated numerically, as shown in Figure~\ref{fig:grad}.  We see that gradients are suppressed by two orders of magnitude relative to the perturbative effect of lensing.

\section{Filters} \label{app:filters}
In this appendix, we will explain how we choose the filters $h$ , $\bar h$ and $g$, which select for, respectively, the CMB modes we delens, the CMB modes we do not delens, and the  $\phi$ modes we include when delensing.  We will first motivate the need for filtering in our procedure for delensing.  We will then determine $\bar h$ in terms of $h$ and $g$ by demanding that the total power is conserved by delensing.  Finally, we discuss how to optimize the choice of $h$ and $g$ to minimize the variance induced in the maps from lensing / delensing and the generalization to the polarization maps.

\subsection{Noise and Filtering}\label{app:noise}

When delensing a temperature or polarization map, it is intuitively clear that we should filter the lensing map used for delensing.  After all, we are trying to remove the effect of the physical lens, not introduce more noise into the maps.  On the other hand, filtering the temperature and polarization maps may not be as obvious.  We often take cross-correlations between different subsets of the data in order to cancel the noise and avoid noise-bias in the resulting spectra.  In this sense, we can remove the noise with filtering.

\begin{figure}[t!]
\begin{center}
\includegraphics[width=0.8\textwidth]{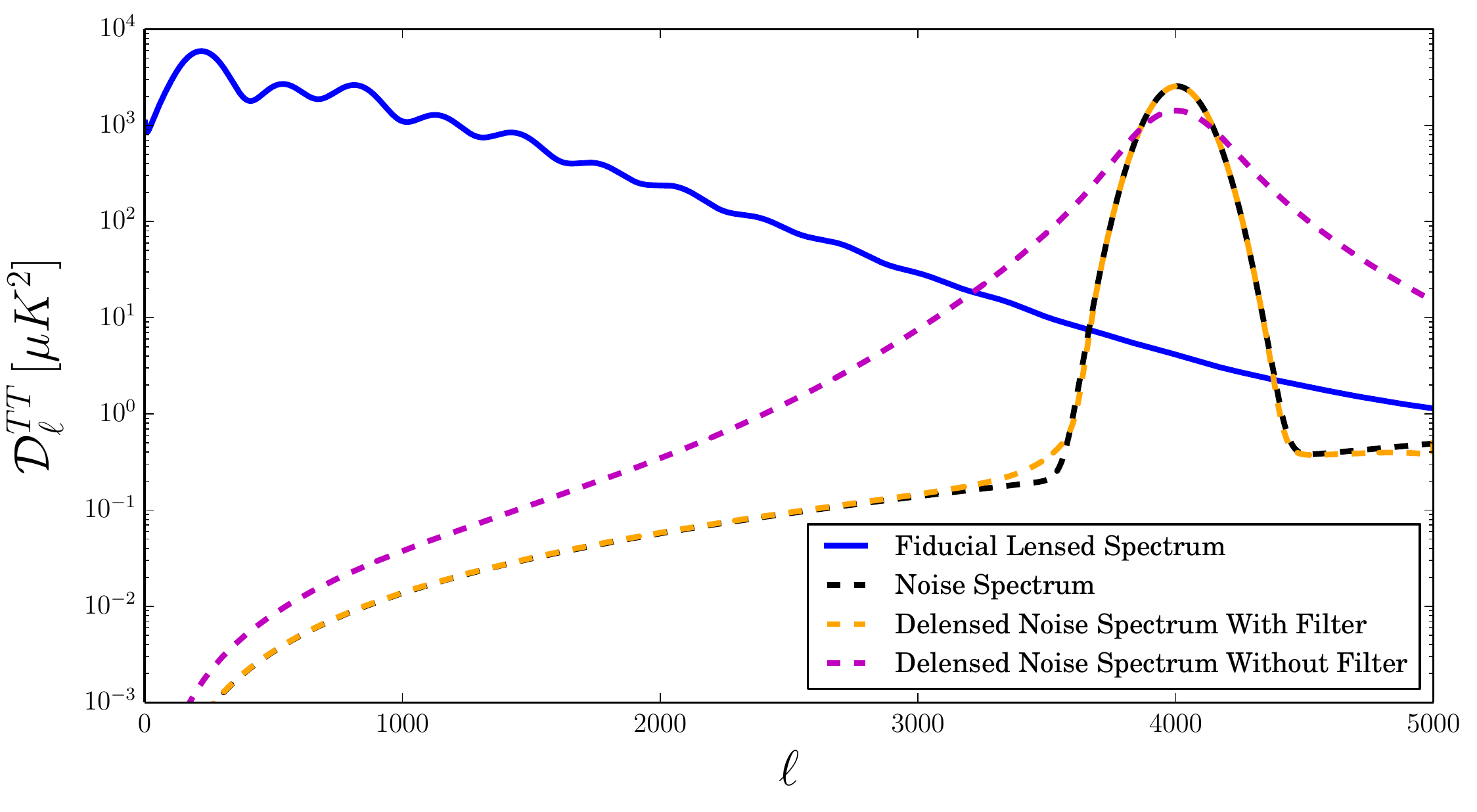}
\caption{Illustrating the value of filtering on noise-dominated CMB modes.  The lensed temperature power spectrum and noise curves in solid blue and dashed black respectively.  The noise spectrum is given a large spike for illustration.  The noise curve after delensing is shown with and without filtering in orange and magenta respectively.  We see that the unfiltered noise curve exceeds the signal in locations that were signal dominated before delensing.  Filtering leaves the noise curve essentially unchanged.  }
\label{fig:effect_of_filtering}
\end{center}
\end{figure}

When it comes to constraining cosmological parameters, the noise will always enter through the covariance matrix.  For example, given a perfect measurement of $\phi$, delensing an unfiltered temperature map ($h=1$, $\bar h = 0$) gives us the map
\beq
\Td(\x) = T(\x) + T^{\rm N}(\x- \va(\x)) \ .
\eeq
While we have removed the lensing from the signal, we have lensed the noise in the process.  The delensed noise power spectrum becomes
\beq
C^{d, N}_\ell = \int d^2 r \frac{d^2 \ell_1 }{(2\pi)^2}  C_{\ell_1}^{\rm N} e^{-i (\vl - \vl_1)\cdot \r} e^{-\frac{\ell_1^2}{2} (\Co_0(0)-\Co_0(r)+ \cos 2\varphi_1\Co_2(r))} \ .
\eeq
Lensing therefore moves the power in $C_\ell^{\rm N}$ to from one scale to another. Most importantly, it will move the noise power from regions with high noise to those with low noise.  As a result, if we do not filter before delensing, we will be allowing the noisiest modes to  corrupt the cleaner modes by moving noise power around.  This effect is illustrated in Figure~\ref{fig:effect_of_filtering}, where we have shown the noise curves before and after delensing with and without filtering.  We have added a region with large non-white noise to show how this noise corrupts modes that were measured with high signal-to-noise before delensing. 

The goal of filtering is therefore to isolate the noisiest modes from the cleanest modes.  There are many ways one could imagine doing this, depending on the goals.  Regardless of motivations, one is always left to find an optimal filtering procedure.  Defining a (near)-optimal choice of filters will become the focus of the rest of this section.   Our goal is not to throw away information but to combine the available maps to produce the best possible measurements of cosmological parameters.

\subsection{Conservation of Total Power} \label{app:totalpowerT}

We want our delensing procedure to conserve total power to mimic the properties of lensing.  Furthermore, conserving power ensures we are not throwing any information away but simply moving it around.  This constraint determines $\bar h$ in terms of $h$ and $g$ from the requirement that
\beq\label{eq:deftotalpower}
\langle \Td(0)^2 \rangle = \langle \To(0)^2 \rangle = \int \frac{d^2 \ell}{(2\pi)^2} (C_\ell + C_\ell^N)\ . 
\eeq
To simplify the calculations, we work with the simplified (dropping gradients) form of delensing
\beq
\Td(\x) \simeq \int \frac{d^2 \ell}{(2\pi)^2} \Big( e^{i \vl \cdot (\x+\va(\x))}\left(\bar h_\ell + e^{-i\vl \cdot g\star \vao(\x)} h_{\ell}\right)  T_{\vl}+e^{i \vl \cdot \x}\left(\bar h_\ell + e^{-i \vl \cdot g\star \vao(\x)} h_{\ell}\right)  T^N_{\vl} \Big)
\eeq
and therefore
\beq
\langle \Td(0)^2 \rangle = \int \frac{d^2 \ell}{(2\pi)^2} \Big( \left(|\bar h_\ell|^2 + |h_\ell|^2\right) C_\ell^{\rm obs} + \left(\bar h_\ell h_{-\ell}+ h_{\ell} \bar h_{-\ell}\right) e^{-\frac{\ell^2}{4} \Co_0(0)} C_{\ell}^{\rm obs}\Big) \ .
\eeq
We solve Equation~(\ref{eq:deftotalpower}) for $\bar h_\ell$ and find
\beq
\bar h_{\ell} = \sqrt{1- h_\ell^2 \left(1-e^{- \frac{\ell^2}{2} \Co_0(0)}\right)} - h_\ell e^{- \frac{\ell^2}{4} \Co_0(0)} \ .  \label{eq:hbar}
\eeq
This formula is not quite correct, due to neglecting gradients of $\va$ but this is a small effect (especially at the level of the filters).  

\subsection{Optimal Filters}

In order to find the optimal result, we need to first define what we are trying to minimize.  In the interest of providing all-orders expressions, we will define our procedure such that we choose $h$ and $g$ to minimize
\beq\label{eqn:minimization}
\left\langle \big( \Td_\vl - \langle \Td_\vl \rangle_{\phi, \phi^{\rm N}}\big) \big(\Td_{-\vl} - \langle \Td_{-\vl} \rangle_{\phi, \phi^{\rm N}}\big) \right\rangle_{T, \phi,\phi^{\rm N} }
\eeq
where $\langle \ldots \rangle_{X}$ means the statistical average with respect to $X$, holding everything else fixed.  The advantage of this choice is that we are minimizing something that is manifestly positive and that it will reproduce the filters used in the perturbative limit.  

The intuition for this choice is as follows: we are demanding that we minimize how much $\Td$ varies with each realization of the lensing field and the reconstruction noise.  In the limit where there is no noise, this means that we have removed the lensing from $\Td$, as the map does not change under different realizations of the lensing potential.  In the limit of a very noisy reconstruction, our procedure minimizes how much noise is introduced into the delensed temperature maps.  As we eventually want to use the delensed maps for cosmological constraints, minimizing the noise in the maps is also a desirable feature.  Finally, we will see in Appendix~\ref{app:info} that this minimization procedure determines an approximate local extremum of the Fisher information, and should be close to providing the best possible limits on cosmological parameters of interest.  

In principle, we can minimize Equation~\ref{eqn:minimization} to determine $g_L$ and $h_\ell$ given any noise levels for $\To$ and $\phi^{\rm obs}$.  Unfortunately, solving these equations in complete generality is challenging, even numerically.  However, the solutions simplify in a number of limits that will allow us to gain  intuition for the behavior of the optimal solution.  In practice, we want filters that are easy to implement on real data and therefore we want a simple filtering scheme that approximates the various limits of the optimal filters.

\vskip 10pt
\noindent {\bf Perturbative Limit:} \hskip 6pt Let us start by expanding $\Td$ to linear order in $\phi$ (ignoring gradients) to get
\bea
\Td_\vl &\simeq & (1- h_\ell) \Big[ \Tn_\vl + T_\vl +\int \frac{d^2 L}{(2\pi)^2} \big(- \vL \cdot (\vl-\vL)\big) \phi_{\vL} T_{\vl-\vL} \Big]  
 \nonumber \\
 && +  h_\ell \Big[ \Tn_\vl+ T_\vl +\int \frac{d^2 L}{(2\pi)^2}\big(- \vL \cdot (\vl-\vL)\big) \big( \phi_{\vL} T_{\vl-\vL} - g_{|\vl-\vl_1|} \phi^{\rm obs}_{\vL} (T_{\vl-\vL}+\Tn_{\vl-\vL})\big) \Big]  \nonumber \\
 &\simeq &  \Tn_\vl + T_\vl  + \int \frac{d^2 L}{(2\pi)^2} \big(- \vL \cdot (\vl-\vL)\big) \big( \phi_{\vL} T_{\vl-\vL}  - h_{|\vl-\vL|} g_{L} \phi^{\rm obs}_{L} (T_{\vl-\vL}+\Tn_{\vl-\vL})\big)
\eea
where we used $\bar h_\ell  = 1- h_\ell + {\cal O}(C^{\phi \phi}_L)$, $\va(\x) = \int \frac{d^2 L}{(2\pi)^2} e^{i \vL \cdot \x} \, i  \vL \phi_{\vL}$ and $h_{\ell} \to h_{|\vl-\vL|}$ (which is the small gradient expansion). We will simplify this expression by assuming $h_{\ell}$ and $g_{L}$ are real (i.e. real in both position- and $\ell$-space).  Now we minimize with respect to $h_{q}$ (noticing that $\langle \Td \rangle_{\phi, \phi^{\rm N} }\simeq  \Tn_\ell + T_\ell $ is independent of $h$) to find
\beq
\partial_h \langle \Td_\vl \Td_{-\vl} \rangle' =- 2 g_{|\vl-\q|} \, C_{q}  C^{\phi\phi}_{|\vl-\q|}  + 2 h_{q} \, g_{|\vl-\q |}^2 \,  ( C_{q}+ C_{q}^{N})  C^{\phi\phi,{\rm obs}}_{|\vl-\q|} = 0 \ ,
\eeq
where $\langle \ldots \rangle'$ means we have removed the $(2\pi)^2 \delta(0)$. To solve this equation, we note that if we write 
\beq
\frac{g_{|\vl  -\q|} C^{\phi\phi,{\rm obs}}_{|\vl-\q |} }{C^{\phi \phi}_{|\vl-\q|} } = \frac{C_{q}}  {h_{q} ( C_{q}+ C_{q}^{N}) }\ ,
\eeq
then the left and right hand sides must be constants (independent of $\vl$ and $\q$), otherwise it would be impossible to find a solution.  This means that 
\bea
g_{\ell } = a \frac{C^{\phi \phi}_{L}}{C^{\phi \phi, {\rm obs}}_{L}} \qquad \qquad h_{\ell} = a^{-1} \frac{C_{\ell}}{C_{\bar \ell}+ C_{\ell}^{N} } \ ,
\eea
where $a$ is a constant.  In the limit of a perfect measurement, we would like $g =h =1$, so we should choose $a=1$.  

Technically speaking, we should have included the order $\phi^2$ term in the expansion of $\Td$ because it will contribute to the power spectrum at this order.  This would contribute a term of the from $\kappa h_\ell T_{\ell} $ where $\kappa \propto \int d^2 L \, L^2 C^{\phi \phi}_L$ is a constant independent of $\ell$.  In what follows, an important feature of this minimization procedure is that if we want to solve $ a(q) b(\ell-q) +c(q) d(\ell) = 0$ (where $a,b,c,d$ are functions of one variable) for all $q, \ell$, then both terms must be independent of $q$ and $\ell$ which means we can look at each term independently.

The take-away from this calculation is that our definition of optimal filtering for delensing matches the perturbative result~\cite{Smith2010} in the appropriate limit.
\vskip 10pt
\noindent {\bf Noisy Lensing Reconstruction:} \hskip 6pt Now suppose we have a noisy measurement of the lensing potential such that the perturbative result suggests we should take $g\ll 1$.  Now we can expand in $g$ as the small number while keeping all orders in $\phi$.  Working to linear order in $g$, we have 
\bea
\Td_\vl &\simeq & (1- h_\ell) \Big[  \Tn_\vl  + \tilde T_\vl \Big] +  h_\ell \Tn_\vl+ h_\ell  \tilde T_\vl +\int \frac{d^2 L}{(2\pi)^2}  \vL \cdot (\vl-\vL)  g_{L} \phi_{\vL}^{\rm obs} h_{|\vl-\vL|}( \tilde T_{\vl-\vL}+\Tn_{\vl -\vL} ) \nonumber \\
&=& \Tn_\vl  + \tilde T_\vl + \int \frac{d^2 L}{(2\pi)^2}  \vL \cdot (\vl-\vL) g_{L} \phi_{\vL}^{\rm obs} h_{|\vl-\vL|}( \tilde T_{\vl-\vL}+\Tn_{\vl -\vL} )
\eea
Now expanding the power spectrum and taking a derivative with respect to $g_{q}$ we get
\beq
2 (\vl-\q) \cdot \q \, h_{|\vl-\q|}  \langle \phi^{\rm obs}_\q \tilde T_{\vl -\q} \tilde T_{-\vl} \rangle' - 2 g_q C_q^{\rm \phi \phi, {\rm obs}}  ((\vl-\q) \cdot \q)^2 |h_{|\vl-\q|}|^2 C^{\rm obs}_{|\vl-\q|} = 0 \ .
\eeq
The statement $g_L \ll 1$ implies that $C^{\phi \phi, N}_L \gg C^{\phi \phi}_L$ and therefore it is consistent to drop terms proportional to $g_L^2 C^{\phi \phi, N}_L$ while keeping terms of order $g_L C^{\phi \phi, N}_L \sim C^{\phi \phi}_L$.  

The last thing we need to evaluate is  $\langle \phi^{\rm obs}_\q \tilde T_{\vl -\q} \tilde T_{-\vl} \rangle$.  Using Gaussian statistics, we see that for a Gaussian random field $y$,
\bea
\langle y \, f(y) \rangle &=& \int dy \frac{1}{\sqrt{2\pi}\sigma} (-\sigma^2) \left(\partial_y e^{-\frac{y^2}{2\sigma^2} }\right) f(y)\nonumber \\
&=& \sigma^2 \langle f'(y)\rangle \ .
\eea
Now we can use (being careful to note that when $y$ is complex we take a derivative with respect to $y^*$),
\bea
\partial_{\phi_{-\q}} \tilde T_{\vl}  &=& \int d^2 x_1 \frac{d^2 \ell_1}{(2\pi)^2} e^{-i \vl \cdot \x_1} (\vl_1\cdot \q) e^{i \q\cdot \x_1} e^{i \vl_1 \cdot  \x_1+\int \frac{d^2 L}{(2\pi)^2}  e^{i \vL \cdot \x_1} (\vl_1\cdot \vL) \phi_\vL} \nonumber \\
&=& (i \q)\cdot (\widetilde{\nabla T})_{\vl + \q} \ ,
\eea
where $\widetilde{\nabla  T} = \nabla T (\x+ \va(\x))$.  Now if we define
\beq
\left\langle (\widetilde{\nabla  T})_\vl \,  \tilde T_{\vlp}\right\rangle = i \vl \tilde C_{\ell}^{T \nabla T} (2\pi)^2 \delta(\vl+\vlp) \ ,
\eeq
then we find
\beq
\langle \phi^{\rm obs}_{\q} \tilde T_{\vl -\q} \tilde T_{-\vl} \rangle' =  (\vl-\q) \cdot \q \, C_q^{\phi \phi} \tilde C^{T \nabla T}_{\ell-q} -\vl \cdot \q  C_q^{\phi \phi} \tilde C^{T \nabla T}_{\ell} \ .
\eeq
Putting this together we find 
\beq
g_{\ell} = \frac{C^{\phi \phi}_L}{C^{\phi \phi, {\rm obs}}_L } \, , \qquad \qquad h_{\ell} = \frac{\tilde C_{\ell}^{T\nabla T}}{C_{\ell}^{\rm obs}} \ .
\eeq
It turns out that $\tilde C_{\ell}^{T\nabla T} \simeq \tilde C_{\ell}$~\cite{Lewis:2011fk} and therefore this explains why taking the perturbative result and making the replacement $C_\ell \to \tilde C_\ell$ is useful approximation to the non-perturbative result, at least in the limit of a noisy lensing map.
\vskip 10pt
\noindent {\bf Ideal Lensing Map:} \hskip 6pt In order to get intuition for the limits of an ideal filter, we will finally consider the case where we have a (nearly) perfect lensing map in the presence of noisy temperature (or polarization) data.

The complication presented by noisy data is that we can perfectly delens the underlying CMB modes, but we will also shift around the noise in the process.  In harmonic space, we will have
\beq
\Td_\vl =\bar h_\ell (\tilde T_\vl + \Tn_\vl) + h_\ell (T_\vl + \tilde{T}^{\rm N}_{\vl})
\eeq
 where $\tilde T^{\rm N}(\x) = T^{\rm N}(\x- \va(\x))$.  We notice that 
 \beq
\Delta \Td_\vl \equiv \Td_\vl -\langle \Td \rangle_{\alpha} = \bar h_\ell( \tilde T_\vl - e^{-\frac{\ell^2}{4} C_0(0)} T_\vl) + h_\ell (\tilde{T}^{\rm N}_{\vl} - e^{-\frac{\ell^2}{4} C_0(0)} T_\vl^{\rm N} )
 \eeq
 so that 
 \bea
 \left\langle \Delta \Td_\vl \Delta \Td_{\vlp} \right\rangle &=& (2\pi)^2 \delta(\vl+\vlp) \Big[ | \bar h_\ell|^2 ( \tilde C_\ell - e^{-\frac{\ell^2}{2} C_0(0)} C_\ell ) + |h_\ell|^2  ( \tilde C_\ell^{\rm N} - e^{-\frac{\ell^2}{2} C_0(0)} C_\ell^{\rm N} ) 
 \eea
 Because of the relatively complicated form of $\bar h_\ell $ in terms of $h_\ell$ this is not especially easy to minimize with respect to $h_\ell$.  However, we are mostly interested in the behavior that defined $h_\ell \ll 1$ so we can expand $\bar h_\ell \simeq 1- h_\ell e^{- \frac{\ell^2}{4} C_0(0)}$.  Minimizing with respect to $h_\ell$ we find
 \beq
 h_\ell \sim  \frac{ e^{- \frac{\ell^2}{4} C_0(0)}\left( \tilde C_\ell - e^{-\frac{\ell^2}{2} C_0(0)} C_\ell\right)}{(\tilde C^{\rm N}_\ell  - e^{-\frac{\ell^2}{2} C_0(0)} C^{\rm N}_\ell)} \ .
 \eeq
 In writing this expression, we assumed that the temperature map is noisy, in order to be consistent with the assumption $h_\ell \ll 1$.  We can evaluate this expression using 
\bea
\Delta \tilde C^{({\rm N} )}_\ell &\equiv&  \left(\tilde C^{({\rm N})}_\ell  - e^{-\frac{\ell^2}{2} C_0(0)} C^{({\rm N})}_\ell\right)  \nonumber \\
&=&  e^{-\frac{\ell^2}{2} C_0(0)}  \int d^2 r \frac{d^2 \ell_1}{(2\pi)^2} e^{-i(\vl-\vl_1)\cdot \r}\left( e^{\frac{\ell_1^2}{2}( C_0(r)- \cos 2\varphi C_2(r))}-1\right) C_{\ell_1}^{({\rm N})} \ .
\eea
The main takeaway is that in the limit of a perfect lensing map, the optimal filter as the temperature map becomes noisy is controlled by the ratio of the lensed power to the delensed noise.  

\vskip 10pt
\noindent {\bf Summary:} \hskip 6pt While our all-orders spectra suggest an all-orders method for choosing filters, in practice the the optimal choice is difficult to determine analytically and therefore of limited practical utility.  However, the all-orders approach simplifies in a number of limits of interest and therefore we can choose our filters by matching the appropriate limit.  For the lensing potential, the optimal fiter in all cases is
\beq
g_L = \frac{C_L^{\phi \phi}}{C_L^{\phi\phi} +C_L^{\phi\phi, {\rm N}} } \ .
\eeq
The case is more complicated for temperature and polarization, as there are three different types of spectra to consider: lensed, unlensed, and delensed spectra.  Intuitively, when we are signal dominated for all of the spectra, the choice $g_L, h_\ell, \hP_\ell \simeq 1$ is free of any subtlety. Furthermore, for $T$ and $E$, as we go to larger $L$ and $\ell$, the noise in $C_L^{\phi \phi, {\rm obs}}$ usually dominates before we are limited by the noise in $T$ or $E$.  Therefore, we will typically be in the situation where $g \ll 1$ when the choice when there is a noticeable difference in the filters for $T$ and $E$.  We calculated the optimal filters perturbatively in $g$ and found 
\beq
\label{eqn:optimal_temp_filter}
h_{\ell} = \frac{\tilde C_{\ell}^{T\nabla T}}{C_{\ell}^{\rm obs}} \simeq \frac{\tilde C_{\ell}^{TT}}{C_{\ell}^{\rm obs}} 
\eeq
This choice seems to behave appropriately in the limits applicable to a typical CMB experiment and would seem to be the appropriate choice for our forecasting purposes.  The corresponding $\bar h$ filter is 
\beq
\bar h_{\ell} = \sqrt{1- h_\ell^2 \left(1-e^{- \frac{\ell^2}{2} \Co_0(0)}\right)} - h_\ell e^{- \frac{\ell^2}{4} \Co_0(0)} \ .
\eeq

\subsection{Polarization}

Having fully explored the choice of filters for temperature, we will now repeat the process for polarization (in an abbreviated form).  Our delensed polarization field is defined to be 
\bea
\big[ \Qd\pm i  \Ud \big] (\x)  =  \bhP \star \big[\Qo\pm i \Uo \big] (\x)  + \hP \star \big[\Qo\pm i \Uo \big] (\x-g\star \vao(\x))  \ .
\eea
As discussed in the main text, we are choosing a common filter for both $Q$ and $U$ and we will take $\bhP$ and $\hP$ to be real.  We choose a common filter because isotropic noise implies the noise is the same for both $Q$ and $U$.  Furthermore, since lensing acts locally on $Q$ and $U$, we want to avoid filtering that mixes $Q$ into $U$ or vise versa.  A direct consequence of these choices is that there is a common filter for $E$ and $B$.  We will expand on this choice in the next subsection.

The first constraint is that we require that the total power is unchanged, which for polarization means we want to keep $Q^2(0)+U^2(0)$ fixed.  This implies that 
\bea
\left\langle \left| \Qd\pm i  \Ud \right|^2 \right\rangle(0)&=&  \int \frac{d^2 \ell}{(2\pi)^2}\Big( C_\ell^{EE, {\rm obs}}+C_\ell^{BB, {\rm obs}} \Big)   \\ 
&& \qquad \qquad \times \left[ \left(\left|\bhP_\ell\right|^2 + \left|\hP_\ell\right|^2\right)  + \left(\bhP_\ell \hP_{-\ell}+ \hP_{\ell} \bhP_{-\ell}\right) e^{-\frac{\ell^2}{4} \Co_0(0)} \right] \nonumber
\eea
Requiring that this is the same as the observed power implies that 
\beq
\bhP_{\ell} = \sqrt{1- {\hP_\ell}^2 \left(1-e^{- \frac{\ell^2}{2} \Co_0(0)}\right)} - \hP_\ell e^{- \frac{\ell^2}{4} \Co_0(0)} \ .
\eeq
Perhaps unsurprisingly, this is identical to the constraint we found for the temperature filters.

Now we will compute the filter in the noisy lens limit by minimizing
\beq
\left\langle \big( \Ed_\vl - \langle \Ed_\vl \rangle_{\phi, \phi^{\rm N}}\big) \big(\Ed_{-\vl} - \langle \Ed_{-\vl} \rangle_{\phi, \phi^{\rm N}}\big) \right\rangle_{E, \phi,\phi^{\rm N} }  \ .
\eeq
Expanding in small $g_\ell$ we find
\bea
\Ed_\vl &=& \Eo_\vl + \int \frac{d^2 L }{(2\pi)^2}\Big[ \hP_{|\vl -\vL|}  \vL \cdot (\vl - \vL) g_{L} \phi^{\rm obs}_{\vL}   \nonumber \\
&&\times \Bigg[  \cos(2 (\varphi_{\vl -\vL} - \varphi_\vl)) \Eo_{\vl-\vL} + \sin(2 (\varphi_{\vl-\vL} - \varphi_\vl))\Bn_{\vl-\vL} \Bigg] \label{eqn:Edfilter}
\eea
The angles introduce a small complication compared to temperature.  However, in practice the lensing is peaked at low multipoles such that $|\vl- \vl_1 | \ll \ell$ or $\cos (\varphi_{\vl_1} -\varphi_{\vl} ) \simeq 1$.  Therefore, to simplify the discussion we can set $(\varphi_{\vl_1} - \varphi_\vl) =0$ such that 
\bea
\langle \Ed_{\vl} \Ed_{\vlp} \rangle &\sim& (2\pi)^2 \delta(\vl +\vlp) C_\ell^{EE, {\rm obs} } +  \int \frac{d^2 L }{(2\pi)^2}\Big[ \hP_{|\vl -\vL|}  \vL \cdot (\vl - \vL) g_{L}  \left\langle \phi^{\rm obs}_{\vL} \tilde E_{\vl-\vL} \tilde E_{\vlp} \right\rangle +\{ \vl \leftrightarrow \vlp \} \Big] \nonumber \\
&&+ \int \frac{d^2 L }{(2\pi)^2}  \left|\hP_{|\vl-\vL|}\right|^2  \left|g_{L}\right|^2   \left[\vL \cdot (\vl - \vL) \right]^2 \, C^{\phi \phi, {\rm obs}}_{L}    C^{EE, {\rm obs}}_{|\vl-\vL|} \ .
\eea
 Now taking a derivative with respect to $g_q$ we get
\beq
\left(  \hP_{|\vl-\q|}  \q \cdot (\vl - \q)   \left\langle \phi^{\rm obs}_{\q} \tilde E_{\vl-\q} \tilde E_{\vlp} \right\rangle +\{ \vl \leftrightarrow \vlp \} \right) = 2\left|\hP_{|\vl-\q|}\right|^2  g_{q}   \left[\q \cdot (\vl - \q) \right]^2 \, C^{\phi \phi, {\rm obs}}_{1}  C^{EE, {\rm obs}}_{|\vl-\q|}  
\eeq
We can evaluate the left hand side using $\langle y f(y) \rangle = \sigma^2 \langle f'(y) \rangle$ for Gaussian $y$ as we did for temperature.  We will define 
\beq
\widetilde{ \nabla E}_\vl \equiv i \left[{\rm Re} \int d^2 x \frac{d^2 \ell_1}{(2\pi)^2}  \, \vl_1 E_{\vl_1} e^{-i (\vl -\vl_1) \cdot \x}  e^{2 i ( \varphi_{\vl_1} - \varphi_{\vl})} e^{i \vl_1\cdot \va(\x)}  \right]
\eeq
and 
\beq
\left\langle \widetilde{ \nabla E}_\vl \tilde E_{\vlp} \right\rangle = i \vl \tilde C^{E\nabla E}_\ell (2\pi)^2 \delta(\vl +\vlp)
\eeq
Following the same procedure as we did for temperature, we have
\bea
\partial_{\phi_{-\q}} \tilde E_{\vl}  &=& (i \q)\cdot ( \widetilde{\nabla E})_{\vl + \vec q} \ .
\eea
and therefore
\beq
\left\langle \phi^{\rm obs}_{\q} \tilde E_{\vl -\q} \tilde T_{\vlp} \right\rangle =  (2\pi)^2 \delta(\vl+\vlp) \Big[(\vl-\q) \cdot \q \, C_q^{\phi \phi} \tilde C^{E \nabla E}_{\ell-q} -\vl \cdot \q \, C_q^{\phi \phi} \tilde C^{E \nabla E}_{\ell} \Big]\ .
\eeq
The optimal filters are then
\beq
\label{eqn:optimal_pol_filter}
g_{L} = \frac{C^{\phi \phi}_L}{C^{\phi \phi, {\rm obs}}_L } \, , \qquad \qquad \hP_{\ell} = \frac{\tilde C_{\ell}^{E\nabla E}}{C_{\ell}^{EE,{\rm obs}}} \approx  \frac{\tilde C_{\ell}^{E E}}{C_{\ell}^{EE,{\rm obs}}}  \ .
\eeq
It is reassuring that $g$ in unchanged from the temperature filters.  One can check that the approximation $\tilde C_{\ell}^{E\nabla E} \approx \tilde C_\ell^{EE}$ is the same one that allowed us to neglect the angular terms in Equation~\ref{eqn:Edfilter}, namely that $\vl -\vL \simeq \vl$.

\subsection{Filtering in $Q/U$ versus $E/B$ } \label{app:EBfilter}

We have chosen to filter locally in terms of the $Q$ and $U$ maps.  Locally, the polarization defines a vector field with independent components and isotropic noise, so this was a natural choice.  However, one could instead imagine converting the map into $E$ and $B$ modes and filtering each separating.  In this subsection, we will explain the relationship between these two approaches in order to further explain the meaning of our choice in filtering $Q$ and $U$.

To simplify our discussion, let us ignore delensing and simply discuss the meaning of the filters directly.  After converting $P(\x)$ into $E_\vl$ and $B_\vl$, it is natural to consider the filtered fields
\beq
E_{\ell}^{f} = h^E_\ell E_\ell \qquad \qquad B_{\ell}^{f} = h^B_\ell B_\ell \ .
\eeq
To make the connection to the original $Q$ and $U$, let us define
\beq
(Q\pm i U)_\vl = \int d^2 x e^{-i \x\cdot \vl} (Q\pm i U)(\x)
\eeq
such that
\bea
e^{\pm i 2 \varphi_\vl} (E \pm i B)_\vl = Q_\vl \pm i U_\vl \ .
\eea
Now if we define the filtered maps as $Q^f_\vl \pm i U^f_\vl \equiv e^{\pm i 2 \varphi_\vl} (E^f \pm i B^f)_\vl $ we have
\bea
Q^f_\vl + i U^f_\vl =  \frac{1}{2} (h^E_\vl + h^B_\vl) (Q+i U)_\vl + \frac{1}{2}(h^E-h^B)_\vl e^{4 i \varphi_\vl} (Q-i U)_\vl 
\eea
One can immediately see that the first term does not mix $Q$ and $U$, while the second term rotates the polarization vector.  

We can now map this back position space in terms of $P(\x) = Q(\x) + i U(\x)$ and $P^f(\x) = Q^f(\x) + i U^f(\x)$ as 
\beq\label{eqn:EBtoP}
P^f (\x)= \int d^2 x' \left[  h^{E+B}(x-x')  P(x') + h^{E-B}(x-x')  P^*(x')\right]
\eeq
where
\bea
h^{E+B}(\x)&\equiv& \int d^2 \ell e^{i \vl \cdot  \x} \frac{1}{2} (h^E+h^B)_\ell  \\
h^{E-B}(\x) &\equiv& \int d^2 \ell e^{i \vl \cdot  \x} e^{i 4 \varphi_\vl} \frac{1}{2} (h^E-h^B)_\ell \ .
\eea
We see that the $h^{E+B}$ filter is an ordinary scalar under rotations while the $h^{E-B}$ transforms like a spin-4 object.

The properties of these two filters are relatively easy to understand from isotropy.   Since we want to isotropy to be preserved by filtering, our filters should also decompose into representations of the rotations.  Furthermore, in order for the filtered field to transform in the same way as the unfiltered field, the filters in Equation (\ref{eqn:EBtoP}) are limited to spin-0 and spin-4.  

Our filtering scheme is equivalent to only the first term in Equation (\ref{eqn:EBtoP}) and therefore we have set $h^{E-B} = 0$ or $h^E = h^B$.  This choice also matches the optimal perturbative result~\cite{Smith2010}.  We made this choice to preserve the local nature of the filter in real space.  Specifically, since $\delta(\x)$ is a scalar under rotations, in the limit where the filter is trivial we can only have the scalar piece.  This is desirable for delensing because delensing itself is local in $\x$.

Although we have chosen to drop the $h^{E-B}$ filter, one could imagine situations where it provides useful information.  Because the primordial $E$-mode signal is much larger than the $B$-mode signal, there are correlations in the signal between $U$ and $Q$ at separated points.  One could imagine this non-local information being useful in weighting the signal-to-noise in the filtered maps, especially for noisy maps.  Whether this improves constraints is a question of whether our filtering scheme is truly optimal.  It is possible that this more elaborate scheme that includes the spin-4 filter could improve constraints; although, as we saw in Section~\ref{sec:params}, there is often very little room for improvement given how close the delensed and unlensed forecasts are when using the simpler filters.

\section{Numeric Computation of Polarization Spectra} \label{app:numpol}

Here we give the expressions used for the numeric computation of delensed polarization spectra.  As for the temperature, we expand to first order in $C_2(r)$ and compute the change to the correlation functions due to lensing and delensing
\bea
\Delta\xi_+^{\rm d}(r) &=& \int \frac{d\ell}{2\pi} \ell (C_\ell^{EE} + C_\ell^{BB}) \Bigg[ -J_0(\ell r) \label{eqn:delta_xiplusd} \\
&& + \left|\bhP_\ell\right|^2 \exp\left[-\frac{\ell^2}{2} (C_0(0)-C_0(r))\right]\left(J_0(\ell r) + \frac{\ell^2}{2} C_2(r) J_2(\ell r)\right)  \nonumber \\
&& + \left(\hP_{\ell} \bhP_{-\ell}+\hP_{-\ell} \bhP_{\ell}\right) \exp\left[-\frac{\ell^2}{2} \left((C_0(0)-C_0(r)) - (\Ccr_0(0)-\Ccr_0(r)) + \frac{1}{2}\Co_0(0)\right)\right] \nonumber \\
&& \quad \times \left(J_0(\ell r) + \frac{\ell^2}{2} \left(C_2(r) - \Ccr_2(r)\right) J_2(\ell r)\right) \nonumber \\
&& + \left|\hP_{\ell}\right|^2 \exp\left[-\frac{\ell^2}{2} \left((C_0(0)-C_0(r)) - 2(\Ccr_0(0)-\Ccr_0(r)) + (\Co_0(0)-\Co_0(r))\right)\right] \nonumber \\
&& \quad \times \left(J_0(\ell r) + \frac{\ell^2}{2} \left(C_2(r) - 2\Ccr_2(r) + \Co_2(r)\right) J_2(\ell r)\right)\Bigg] \ , \nonumber
\eea
\bea
\Delta\xi_-^{\rm d}(r) &=& \int \frac{d\ell}{2\pi} \ell (C_\ell^{EE} - C_\ell^{BB}) \Bigg[ -J_4(\ell r)  \label{eqn:delta_ximinusd} \\
&& + \left|\bhP_\ell\right|^2 \exp\left[-\frac{\ell^2}{2} (C_0(0)-C_0(r))\right]\left(J_4(\ell r) + \frac{\ell^2}{4} C_2(r) (J_2(\ell r) + J_6(\ell r) )\right)  \nonumber \\
&& + \left(\hP_{\ell} \bhP_{-\ell}+\hP_{-\ell} \bhP_{\ell}\right) \exp\left[-\frac{\ell^2}{2} \left((C_0(0)-C_0(r)) - (\Ccr_0(0)-\Ccr_0(r)) + \frac{1}{2}\Co_0(0)\right)\right] \nonumber \\
&& \quad \times \left(J_4(\ell r) + \frac{\ell^2}{4} \left(C_2(r) - \Ccr_2(r)\right) (J_2(\ell r) + J_6(\ell r))\right) \nonumber \\
&& + \left|\hP_{\ell}\right|^2 \exp\left[-\frac{\ell^2}{2} \left((C_0(0)-C_0(r)) - 2(\Ccr_0(0)-\Ccr_0(r)) + (\Co_0(0)-\Co_0(r))\right)\right] \nonumber \\
&& \quad \times \left(J_4(\ell r) + \frac{\ell^2}{4} \left(C_2(r) - 2\Ccr_2(r) + \Co_2(r)\right) (J_2(\ell r) + J_6(\ell r))\right)\Bigg] \ , \nonumber
\eea
\bea
\Delta\xi_X^{\rm d}(r) &=& \int \frac{d\ell}{2\pi} \ell C_\ell^{TE} \Bigg[ -J_2(\ell r) \label{eqn:delta_xiXd} \\
&& + \bar h_{\ell} \bhP_{-\ell} \exp\left[-\frac{\ell^2}{2} (C_0(0)-C_0(r))\right]\left(J_2(\ell r) + \frac{\ell^2}{4} C_2(r) (J_0(\ell r) + J_4(\ell r) )\right)  \nonumber \\
&& + \left(h_{\ell} \bhP_{-\ell}+\hP_{-\ell} \bar h_{\ell}\right) \exp\left[-\frac{\ell^2}{2} \left((C_0(0)-C_0(r)) - (\Ccr_0(0)-\Ccr_0(r)) + \frac{1}{2}\Co_0(0)\right)\right] \nonumber \\
&& \quad \times \left(J_2(\ell r) + \frac{\ell^2}{4} \left(C_2(r) - \Ccr_2(r)\right) (J_0(\ell r) + J_4(\ell r))\right) \nonumber \\
&& + h_{\ell} \hP_{-\ell} \exp\left[-\frac{\ell^2}{2} \left((C_0(0)-C_0(r)) - 2(\Ccr_0(0)-\Ccr_0(r)) + (\Co_0(0)-\Co_0(r))\right)\right] \nonumber \\
&& \quad \times \left(J_2(\ell r) + \frac{\ell^2}{4} \left(C_2(r) - 2\Ccr_2(r) + \Co_2(r)\right) (J_0(\ell r) + J_4(\ell r))\right)\Bigg] \ . \nonumber
\eea
The delensed polarization spectra are then
\bea
C_\ell^{{\rm d}, EE} &=& C_\ell^{EE} + 2\pi \int rdr \frac{1}{2} \left(J_0(\ell r) \Delta\xi_+^{\rm d}(r) + J_4(\ell r) \Delta\xi_-^{\rm d}(r) \right) \, ,  \label{eqn:delensed_pol_spec} \\
C_\ell^{{\rm d}, BB} &=& C_\ell^{BB} + 2\pi \int rdr \frac{1}{2} \left(J_0(\ell r) \Delta\xi_+^{\rm d}(r) - J_4(\ell r) \Delta\xi_-^{\rm d}(r) \right) \, ,  \nonumber \\
C_\ell^{{\rm d}, TE} &=& C_\ell^{TE} + 2\pi \int rdr J_2(\ell r)  \Delta\xi_X^{\rm d}(r) \, . \nonumber
\eea 

\section{Calculating the Covariance} \label{app:covmat}

In our forecasts, we use the approximate form of the covariance matrix
\bea\label{eqn:appCov}
{\rm Cov}_{\ell \, \ell'}^{XY,WZ} &=& \frac{1}{2 \ell+1} \left[ C_{\ell}^{XW} C_{\ell}^{YZ} + C_{\ell}^{XZ} C_{\ell}^{YW} \right] \delta_{\ell \,\ell'} +\sum_\ml \Bigg[ \frac{\partial C_{\ell}^{XY}}{\partial C_\ml^{\phi \phi}} {\rm Cov}_{\ml \, \ml}^{\phi \phi, \phi \phi} \frac{\partial C_{\ell'}^{WZ}}{\partial C_\ml^{\phi \phi}}\Bigg] \ ,
\eea
where $XY, WZ = TT, TE, EE$ and $C^{XY, WZ}_\ell$ are the delensed spectra.  In order to evaluate the covariance, we therefore need to compute the derivatives $\frac{\partial C_{\ell}^{XY}}{\partial C_L^{\phi \phi}}$.  As a warm up, consider this derivative acting on the lensed $TT$ power spectrum,
\beq
\tilde C_{\ell}^{TT} = \int d^2 r \frac{d^2 \ell_1 }{(2\pi)^2}   C_{\ell_1} e^{-i (\vl - \vl_1) \cdot \r} e^{-\frac{\ell^2_1}{2} (C_0(0)-C_0(r)+ \cos 2\varphi_1 C_2(r))} 
\eeq
such that
\beq
\frac{\partial \tilde C_{\ell}^{TT}}{\partial C^{\phi \phi}_\ml } =   \int d^2 r \frac{d^2 \ell_1 }{(2\pi)^2}   C_{\ell_1} \frac{\ell^2_1 \ml^3}{2}\bigg(J_0(\ml r) -1  - \cos 2\varphi_1 J_2(\ml r) \bigg) e^{-i (\vl - \vl_1) \cdot \r} e^{-\frac{\ell^2_1}{2} (C_0(0)-C_0(r)+ \cos 2\varphi_1 C_2(r))}  \ .
\eeq
Now we can the same procedure to the delensed spectra, holding the filters fixed.  We find 
\bea\label{eqn:Cphideriv}
\frac{\partial C_{\ell}^{{\rm d},TT}}{\partial C^{\phi \phi}_\ml } &=& |\bar h_{\ell}|^2 \frac{\partial \tilde C_{\ell}^{TT}}{\partial C^{\phi \phi}_\ml }+ ( h_{\ell} \bar h_{-\ell}+h_{-\ell} \bar h_{\ell}) (-\tfrac{1}{4}g_\ml^2   \ell^2 \ml^3 ) e^{-\tfrac{1}{4} \ell^2 \Co_0(0)} C_{\ell}^{\rm N} \\
&&+  \int d^2 r \frac{d^2 \ell_1 }{(2\pi)^2} |h_{\ell_1}|^2 C_{\ell_1}^{\rm N} e^{-i (\vl - \vl_1)\cdot \r} e^{-\frac{\ell_1^2}{2} (\Co_0(0)-\Co_0(r)+ \cos 2\varphi_1\Co_2(r))} \nonumber \\
&& \qquad \qquad \times g_\ml^2 \frac{\ell^2_1 \ml^3}{2}\bigg(J_0(\ml r) -1  - \cos 2\varphi_1 J_2(\ml r) \bigg) \nonumber \\
&&+ \int d^2 r \frac{d^2 \ell_1 }{(2\pi)^2}   C_{\ell_1} e^{-i (\vl - \vl_1)\cdot \r} e^{-\frac{\ell^2_1}{2} (C_0(0)-C_0(r)+ \cos 2\varphi_1 C_2(r))} \nonumber \\
&&\qquad \times \frac{\ell^2_1 \ml^3}{2}\bigg[ \Big(J_0(\ml r) -1  - \cos 2\varphi_1 J_2(\ml r)\Big)(1- g_\ml) -\frac{1}{2} \bigg]\nonumber \\
&&\times \Bigg( (h_{\ell_1} \bar h_{-\ell_1}+h_{-\ell_1} \bar h_{\ell_1}) e^{-\frac{\ell_1^2}{4} \Co_0(0) +\frac{\ell_1^2}{2} (\Ccr_0(0)  - \Ccr_0(r))+\tfrac{\ell_1^2}{2}  \cos2 \varphi_{1} \Ccr_2(r))}\Bigg)  \nonumber \\
&&+ \int d^2 r \frac{d^2 \ell_1 }{(2\pi)^2}   C_{\ell_1} e^{-i (\vl - \vl_1)\cdot \r} e^{-\frac{\ell^2_1}{2} (C_0(0)-C_0(r)+ \cos 2\varphi_1 C_2(r))} \nonumber \\
&&\qquad \times \frac{\ell^2_1 \ml^3}{2}\bigg[ \Big(J_0(\ml r) -1  - \cos 2\varphi_1 J_2(\ml r)\Big)(1- g_\ml)^2  \bigg]\nonumber \nonumber \\
&&\times \Bigg( |h_{\ell_1}|^2 e^{-\frac{\ell_1^2}{2} (\Co_0(0)-\Co_0(r) + \cos 2\varphi_1 \Co_2(r) )+\ell_1^2(\Ccr_0(0) - \Ccr_0(r))+  \ell_1^2 \cos2 \varphi_{1}  \Ccr_2(r)} \Bigg) \nonumber 
\eea
In practice, we will always consider the case where the spectra involved are cross-correlations of the from $X = W = T$ and $Y=Z=T'$ such that $C^{{\rm N}, TT'}_\ell = 0$.  Since the off-diagonal terms in Equation~(\ref{eqn:appCov}) will involve only these cross-correlations, we may drop the noise terms in Equation~(\ref{eqn:Cphideriv}) and similarly for covariances including the E-modes.  

\section{Fisher Information and Delensing} \label{app:info}

From some perspectives, it is not clear that errors bars should improve by delensing the temperature and/or polarization.  Cosmological parameters have an impact on lensing and one might worry that delensing could remove this information.  Alternatively, one might argue that all of the cosmological information is in the lensing power spectrum and this information is available whether or not we delens, as long as we include the lensing likelihood, and there should be no improvement by delensing.   Neither of these arguments would suggest that delensing is adding information.  The purpose of this appendix is to address these concerns and show that delensing should always add Fisher Information and reduce error bars (or at least leave them unchanged).

We will assume that $\phi$ is a Gaussian random field, as we did in the main text.  As a result, all information about cosmological parameters encoded in $\phi$ is determined by $C^{\phi \phi}_L$.  Although it does not carry cosmological information itself, this does not imply that the the specific realization of $\phi$ cannot impact the measurement of cosmological parameters.  The realization of $\phi$ lenses the CMB in a way that changes the sensitivity of the lensed spectra to cosmological parameters\footnote{This is most obvious in the case of $r$, where lensing $B$ modes act as a foreground.  For sufficiently low $r$, without delensing with the realization of $\phi$, we could not distinguish $r$ from cosmic variance of the lensing potential. }.  Measuring $\phi$ and removing its effects from the CMB maps can increase the Fisher information by restoring information that was originally in the unlensed spectra. 

We will first demonstrate that our delensing procedure produces a local extremum of the Fisher information.  The easiest way to see this is to note that $\bhP = \bar h = 1$ and $\hP = h = 0$ is equivalent to not delensing.  Therefore, any statement about Fisher information with or without delensing is equivalent to a statement about the choice of filters.  The Fisher matrix for a set of cosmological parameters $\lambda_i$ is given by
\bea\label{eqn:fisher}
F_{ij} = \sum_{X,Y,W,Z} \frac{\partial  C_\ell^{{\rm d}, XY}}{\partial \lambda_i} {\rm Cov}^{-1}_{XY, WZ; \ell, \ell'}  \frac{\partial  C_{\ell'}^{{\rm d}, WZ}}{\partial \lambda_j} 
\eea
where $X,Y,W,Z \supset \{ T, E, B, \phi \}$ and $C^{{\rm d}, \phi\phi }_\ell \equiv C^{\phi \phi}_L$.  We will assume the covariance matrix is given in terms of $C^{{\rm d}, XY}_\ell$ as explained in Appendix~\ref{app:covmat}.  We can determine if delensing will improve the constraint on a given a cosmological parameter by computing
\beq
\frac{\partial}{\partial {\bf h}} F_{ii}|_{{\bf h}=0}  \approx \sum_{X,Y,W,Z} 2 \left( \left. \frac{\partial^2  C_\ell^{{\rm d}, XY}}{\partial \lambda_i \partial {\bf h}} \right|_{{\bf h } = 0}  \right) \, {\rm Cov}^{-1}_{XY, WZ; \ell, \ell'}  \frac{\partial  C_{\ell'}^{{\rm d}, WZ}}{\partial \lambda_i} 
\eeq
where ${\bf h} = \{ h_\ell, \hP_{\ell} \}$.  We have assumed that the change to the Fisher information in $\lambda_i$ is dominated by the change to $C_\ell^{XY}$ rather than to the covariance matrix.  Now we compute
\beq	
\left. \frac{\partial^2  C_\ell^{{\rm d}, XY}}{\partial \lambda_i \partial {\bf h}} \right|_{{\bf h } = 0} = \frac{\partial}{\partial \lambda_i}\Big( \left\langle X_{\vl}^{\rm obs} Y^{\rm obs}(\x - g\star\vao(\x))_{-\vl} \right\rangle'  - e^{- \frac{\ell^2}{4} \Co_0(0)} C_\ell^{XY, {\rm obs}} \Big)+ \{ X \leftrightarrow Y \}
\eeq
where $X,Y \supset \{T,E,B \}$.  Note that $e^{- \frac{\ell^2}{4} \Co_0(0)} C_\ell^{{\rm d}, XY}$ is the change to $C^{XY}_{\ell}$ by the {\it average} effect of lensing on $X$- and $Y$-maps.  The term in brackets is therefore the effect of removing the realization of $\phi^{\rm obs}$ from the maps (i.e.~the total minus the average).  It is easy to see that this does not vanish and will depend on the unlensed spectra.  Therefore, any cosmological parameter that affects the unlensed spectra should have $\frac{\partial}{\partial {\bf h}} F_{ii}|_{{\bf h}=0} \neq 0$.  As a result, we are guaranteed that delensing changes the Fisher information for some of the cosmological parameters (since $h$ can take either sign, we can always increase the Fisher information and therefore decrease the error bars).

We also have good reason to think that our filters are nearly maximizing the Fisher information.  We can see that Equation~\ref{eqn:fisher} is a function only of the delensed spectra.  As a result $\partial_{\bf h} F_{ij} \propto \partial_{\bf h} C^{\rm{d}, XY}$.  One can check that our optimal filters, defined by minimizing Equation~\ref{eqn:minimization}, satisfy
\beq
\partial_{\bf h} C^{\rm{d}, XX}  = 0 \to \partial_{\bf h} F_{ij} \approx 0 \ .
\eeq
In the last step we have assumed that the diagonal terms in the delensed covariance play the dominant role in the Fisher information.  Under these circumstances, optimizing our filters is roughly the same as extremizing the Fisher information.  This is confirmed through our forecasts in Section~\ref{sec:params} where we see that for parameters like $\Neff$, $Y_p$, and $\theta_s$ that affect the primary CMB, the off-diagonal covariances have essentially no effect on the forecasts.

The question that remains is whether the local extremum is a minimum or a maximum.  For parameters that affect only the unlensed spectra, we should maximize information with perfect delensing and therefore this extremum should be a maximum even for imperfect delensing.  On the other hand, it is less obvious that this is a maximum for cosmological parameters that affect $C_{L}^{\phi\phi}$ without introducing large effects in the unlensed CMB.  In such cases, it would seem surprising that delensing increases the Fisher information as we are removing information about the lensing potential from the spectra.  The intuitive reason that delensing does not remove information is that the information that allows us to delens also gives a direct measurement of $C^{\phi\phi}_L$.  For a Gaussian random field, the power spectrum should contain all of the cosmological information encoded in $\phi$.  Therefore, as long as $C^{\phi\phi, {\rm obs}}_L$ is included in the likelihood, we should not gain or lose information encoded in $C_{L}^{\phi\phi}$ by delensing the other spectra.  

To see this this another way, there is also nothing that forbids us form increasing the amount of lensing by changing the sign of $g_L$.  Therefore, if we lost information by delensing, then we should increase information by increasing the amount of lensing in the spectra.  Any such procedure is just some operation performed on a Gaussian random field, $f(\ao)$.  As long as we are only interested information that is contained directly in $C^{\phi\phi, {\rm obs}}_L$, then adding $f(\ao)$ is just repeating the same information and should not be double counted.  Of course, it is the covariance matrix that should correct for this.

To see the role of the covariance for delensing, we will assume that we can measure $\va$ without noise.  The first case we consider is where $C^{\phi \phi}_L$ is not included in our likelihood and we include only the delensed $T,E$, and $B$.  When delensing can be performed perfectly, we have
\beq	
\left. \frac{\partial^2  C_\ell^{{\rm d}, XY}}{\partial \lambda_i \partial {\bf h}} \right|_{{\bf h } = 0} = \frac{\partial}{\partial \lambda_i}\Big(  e^{- \frac{\ell^2}{4} C_0(0)}  C_\ell^{XY}  - e^{- \frac{\ell^2}{4} C_0(0)} \tilde C_\ell^{XY} \Big)+ \{ X \leftrightarrow Y \} \ .
\eeq
Since $C_\ell - \tilde C_{\ell} \sim {\cal O}(C^{\phi\phi}_L)$, we can ignore the derivatives that act on the exponents (to first approximation) and therefore if $\partial_{\lambda_i}C^{XY}_\ell = 0$ we have
\bea
\frac{\partial}{\partial {\bf h}} F_{ii}|_{{\bf h}=0}  &\approx&  \sum_{X,Y,W,Z}  - 2 e^{- \frac{\ell^2}{4} C_0(0)}  \left( \left. \frac{\partial  \tilde C_\ell^{XY}}{\partial \lambda_i \partial {\bf h}} \right|_{{\bf h } = 0}  \right) \, {\rm Cov}^{-1}_{XY, WZ; \ell, \ell'}  \frac{\partial  \tilde C_{\ell'}^{ WZ}}{\partial \lambda_i} \nonumber \\
 &\approx& -2 e^{- \frac{\ell^2}{4} C_0(0)} F_{ii}  < 0
\eea
This is intuitively clear, we are removing the information about $C^{\phi\phi}_L$ and therefore the information decreases.

To see that no information is lost when we include the measurement of $C^{\phi \phi}_{L}$ we must account for the off-diagonal contributions to the covariance matrix.  Inverting the covariance matrix is of course very numerically challenging in practice, but we can schematically understand the effect as
\beq
{\rm Cov}^{-1}_{XX, \phi\phi} \approx - {\rm Cov}_{XX,XX}^{-1}  \frac{\partial C^{{\rm d},XX}}{\partial C^{\phi \phi}} 
\eeq
which is inspired by Equation~\ref{eqn:cov} and inverted as a $2\times2$ matrix assuming we are looking at a single additional spectrum $C^{{\rm d}, XX}$.  We also assumed the off-diagonal covariances are small compared to the diagonal terms.  The final step is to notice that if the only dependence on $\lambda_i$ is through lensing, then
\beq
\frac{\partial C^{{\rm d}, XY}}{\partial \lambda_i} = \frac{\partial C^{{\rm d}, XY}}{\partial C^{\phi\phi}} \frac{\partial C^{\phi\phi}}{\partial \lambda_i}
\eeq
Now combining the off-diagonal and diagonal contributions, 
\bea
 F_{ii} &\approx&  \frac{\partial C^{\phi \phi}}{\partial \lambda_i} {\rm Cov}_{\phi\phi,\phi\phi}^{-1} \frac{\partial C^{\phi \phi}}{\partial \lambda_i} + \sum_{X} \frac{\partial C^{\phi \phi}}{\partial \lambda_i} \frac{\partial C^{XX}}{ \partial C^{\phi \phi}} {\rm Cov}_{XX,XX}^{-1} \frac{\partial C^{XX}}{ \partial C^{\phi \phi}}\frac{\partial C^{\phi \phi}}{\partial \lambda_i} \nonumber \\
 &&\qquad\qquad - \frac{\partial C^{\phi \phi}}{\partial \lambda_i} \frac{\partial C^{XX}}{ \partial C^{\phi \phi}} {\rm Cov}_{XX,XX}^{-1}  \frac{\partial C^{XX}}{\partial C^{\phi \phi}} \frac{\partial C^{\phi \phi}}{\partial \lambda_i}  \nonumber \\
 &\approx& \frac{\partial C^{\phi \phi}}{\partial \lambda_i} {\rm Cov}_{\phi\phi,\phi\phi}^{-1} \frac{\partial C^{\phi \phi}}{\partial \lambda_i} 
 \eea
 This was a vast oversimplification of the problem, but we see that schematically, the purpose of the off-diagonal terms in the covariance is to remove the information that is already included in $C^{\phi\phi}_\ell$, which is all of the non-trivial information in this case.  Therefore, we only see a decrease in the information when we delens if we neglect to include the observed lensing power spectrum.
 
The additional complication of real data (beyond inverting large matrices) is that we do not measure $\phi$ perfectly, and there can be residual information about $C^{\phi \phi}_{L}$ left in the delensed spectrum.  For example, suppose we use a sub-optimal measurement of $\phi$ and therefore both the delensing procedure and $C^{\phi \phi, {\rm obs}}_{L}$ miss important information that is in the lensed spectra.  This information is still encoded in the temperature and polarization spectra, but delensing using $\phi^{\rm obs}$ may decrease the Fisher information by making it more difficult to extract this information.  Furthermore, if we did not filter the noisy modes in $\phi$, then we could imagine the induced error in the delensed spectra could dominate over the error in the observed spectra.

In practice, most information about $C^{\phi\phi}_L$ can be determined through known procedures for extracting $\vao$.  We expect that any information about lensing that is encoded in the map should allow for a reconstruction of $\va$ with low noise.  Therefore, while it is possible for the Fisher information to decrease with delensing, we expect this to occur only when sub-optimal methods are used for reconstruction and/or filtering.

\section{Exact Delensing and Real Data}\label{app:exact}

Given an observed temperature map, $\To(\x)$, and an observed lensing map $\vao$, an alternate approach to delensing is to define \AVE{\cite{anderes14}}
\beq\label{eqn:JobsTd}
\Td(\x) = \bar h \star \To(\x) + \int d^2 x' J^{\rm obs}(\x') \delta(\x - g\star \vao(\x') -\x') h\star \To(\x') \ ,
\eeq
where $J^{\rm obs} = \det \partial_i (x_j + g\star \ao_j )$.  As discussed in Section~\ref{sec:theory}, in the absence of noise with $\bar h=0$ and $h_\ell = g_\ell = 1$, this choice of delensing has $\Td(\x) = T(\x)$ even including gradients.

Assuming the noise is uncorrelated with the lensing noise and the signal, we can formally write the delensed $C_\ell$ as
\bea
\langle \Td_\vl \Td_{\vlp} \rangle &=& (2\pi)^2 \delta(\vl+\vlp) |\bar h_\ell|^2 C_\ell^{\rm obs}+\Bigg[ \int d^2 x'  \frac{d^2 x_1 d^2 \ell_1 d^2 x_2}{(2\pi)^2}  e^{i \vl_1 \cdot (\xp-\x_1) } e^{-i \vl \cdot \xp} e^{-i \vlp \cdot \x_2} \times \nonumber \\
&& \qquad \qquad  \bigg(  \bar h_{\ell'} h_{\ell_1} \left\langle J^{\rm obs}(\x') e^{-i \vl \cdot g\star \vao(\x')} \To(\x_1) \To(\x_2) \right\rangle \bigg) + \{ \vl \leftrightarrow \vlp \} \Bigg] \nonumber \\
&& + \int d^2 x'  d^2 x'' \frac{d^2 x_1 d^2 \ell_1 d^2 x_2 d^2 \ell_2}{(2\pi)^4} e^{-i \vl \cdot \xp -i \vlp \cdot \xp'} e^{i \vl_1 \cdot (\xp-\x_1) } h_{\ell_1} h_{\ell_2 } \times \nonumber  \\
&& \qquad  \qquad  \left\langle J^{\rm obs}(\x') J^{\rm obs}(\x'') e^{-i \vl \cdot g\star \vao(\x') -i \vlp \cdot g\star \vao(\x'') }  \To(\x_1) \To(\x_2)\right\rangle \label{eqn:noapproxjacobian} \ .
\eea
This result can be used as the starting point for a variety of calculations.  In principle, this defines the all-orders result, if one can evaluate all of the correlation functions exactly and perform the integrals.  One may also use this as a starting point where one can systematically include small effects as a perturbative expansion.

There are two main technical challenges of this approach: (1) computing $J^{\rm obs}(\x)$ with real data and (2) evaluating the integral in Equation~\ref{eqn:JobsTd}.  As we saw in Appendix~\ref{app:grad}, including $J^{\rm obs}(\x)$ is necessary for controlling the effects of small scale gradients.  The advantage of our approximate approach in Equation~\ref{eqn:Tddefinition} is that we avoid both complications by simply moving the points in real space.

\clearpage
\phantomsection
\addcontentsline{toc}{section}{References}
\bibliographystyle{utphys}
\bibliography{Refs}

\end{document}